\documentclass{PoS}

\title{DIS2015 Heavy Flavours Working Group Summary}

\ShortTitle{Heavy-flavour Physics Highlights}

\rightline{MAN/HEP/2015/16}

\author{\speaker{Marco Guzzi}\\
        Consortium for Fundamental Physics, School of Physics \& Astronomy\\
University of Manchester, Manchester, M13 9PL, United Kingdom\\
        E-mail: \email{marco.guzzi@manchester.ac.uk}}

\author{Achim Geiser\\
        Deutsches Elektronen-Synchrotron DESY,\\
        Notkestrasse 85, 22607 Hamburg, Germany \\
        E-mail: \email{achim.geiser@desy.de}}

\author{Flera Rizatdinova\\
        Department of Physics, Oklahoma State University,\\ 
        145 Physical Sciences Bldg., Stillwater, OK, USA\\
        E-mail: \email{Flera.Rizatdinova@cern.ch}}

\abstract{Studies presented in the heavy flavours working group are summarized. 
Very recent results of measurements at the HERA, LHC, Tevatron, STAR, PHENIX, and BaBar experiments are reviewed 
and new developments in theory and phenomenology are discussed. In particular, aspects of the 
impact of heavy flavours on global QCD analyses to determine the structure of the proton, 
and analyses in physics beyond the Standard Model are considered.}

\FullConference{The XXIII International Workshop on Deep Inelastic Scattering and Related Subjects\\
		April 27 - May 1, 2015\\
                Southern Methodist University\\
		Dallas, Texas 75275}

\begin{document}

\section{Introduction}
Heavy-flavor physics plays a leading role in many areas of collider phenomenology 
and provides us with powerful tools to handle perturbative and non-perturbative aspects of QCD.

Experimental measurements of cross sections for the production 
of heavy quarks and different charmed and beauty hadrons,
are extremely important to constrain both the 
structure of the proton in various kinematic regions and 
fundamental parameters of the Standard Model (SM), such as quark masses and couplings.
Precision in the determination of the parton distribution functions (PDFs) of the proton 
is crucial for the interpretation of measurements in hadronic collisions. 
PDFs represent a limiting factor in the accuracy 
of theory predictions for SM processes at the LHC.
In this workshop the H1, ZEUS, Tevatron, ATLAS, CMS, and LHCb experiments presented 
novel measurements which are of high importance
for the determination of unpolarized PDFs.
Moreover, it has been reported about the progress going on 
on the theoretical side to reach the new {\it state-of-the-art} in high-order 
QCD calculations to determine fully differential cross sections 
for heavy-flavour production at hadron colliders. 
Heavy-flavour production (in particular production of top-quark pairs $t\bar{t}$) 
has the potential to impose clean constraints on PDFs (gluon) at large $x$, 
where they are currently weak.

Heavy-flavour decays were also extensively discussed, in particular new measurements for 
the B-meson decay and rare charm and beauty-quark decays were presented by the ATLAS, LHCb, and Babar collaborations.
Heavy-flavour decays are important to understand the flavour structure of the SM.   
At the present time, the SM leaves many open questions 
about the flavour sector: the origin of generation and masses, the mixing and disappearance of antimatter. 
Investigations on charge and parity (CP) violations in hadronic final states and on rare decays 
of hadrons made of heavy flavours, offer a powerful way to set constraints on new physics.
Furthermore, heavy quarks hadronize in various charmed and beauty hadrons
with a large number of possible decays to different final states. 
This increases the observability of CP violation effects.
Therefore, precision measurements of heavy-flavour decays provide us with stringent 
tests of the Cabibbo-Kobayashi-Maskawa (CKM) theory~\cite{Cabibbo:1963yz,Kobayashi:1973fv}.

A session of the workshop was dedicated to heavy-flavour production in heavy-ion collisions. 
Heavy-flavour production in high-energy collisions of heavy ions and protons is an excellent probe of the 
quark-gluon plasma (QGP) state of matter. Investigations of extreme conditions of QCD matter 
at very high density and temperature open a way to understand the universe in a few millionths 
of a second after the Big Bang, and to set strong constraints 
on various model calculations for heavy-flavour QGP interactions.  
New heavy-flavour production measurements in proton and deuteron heavy-ion collisions   
were presented by the PHENIX, STAR, and ALICE collaborations.

Many future high-energy physics programs at colliders for the next years will be 
focussed on the search for any possible signature of physics beyond the SM (BSM). 
All of the topics presented in the heavy-flavour working group sessions of the DIS2015 workshop 
are extremely important in this respect. Heavy-flavor physics will play a crucial role in achieving this goal. 
This was reflected in all experimental and theoretical analyses presented, 
which triggered a large number of interesting and lively discussions.
In what follows, the studies presented in the heavy flavour working group 
sessions are briefly summarized in chronological order.

\section{Heavy-flavour production in heavy-ion collisions}

\begin{figure}[ht]
\begin{center}
\includegraphics[width=5cm]{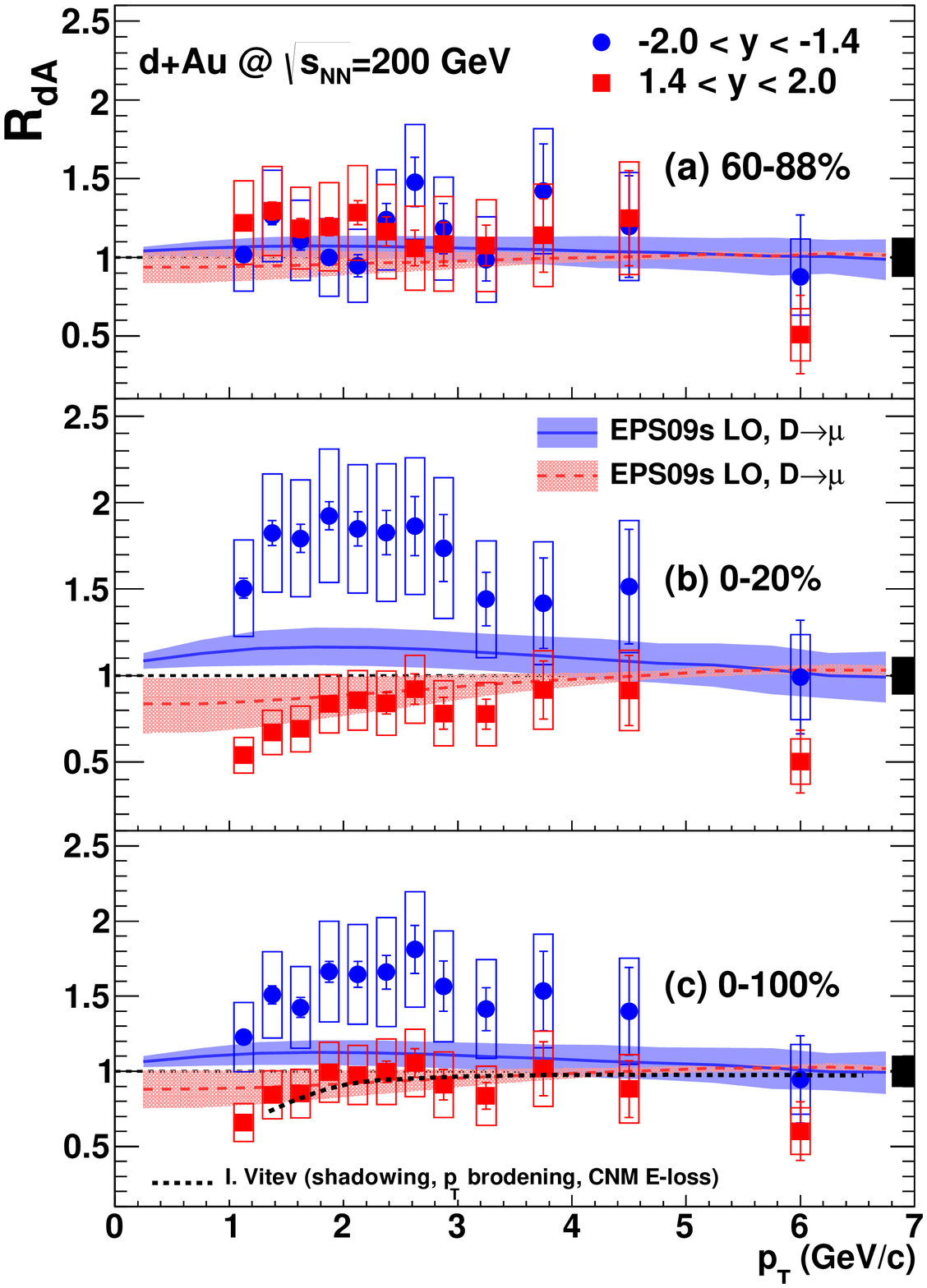}
\includegraphics[width=7cm]{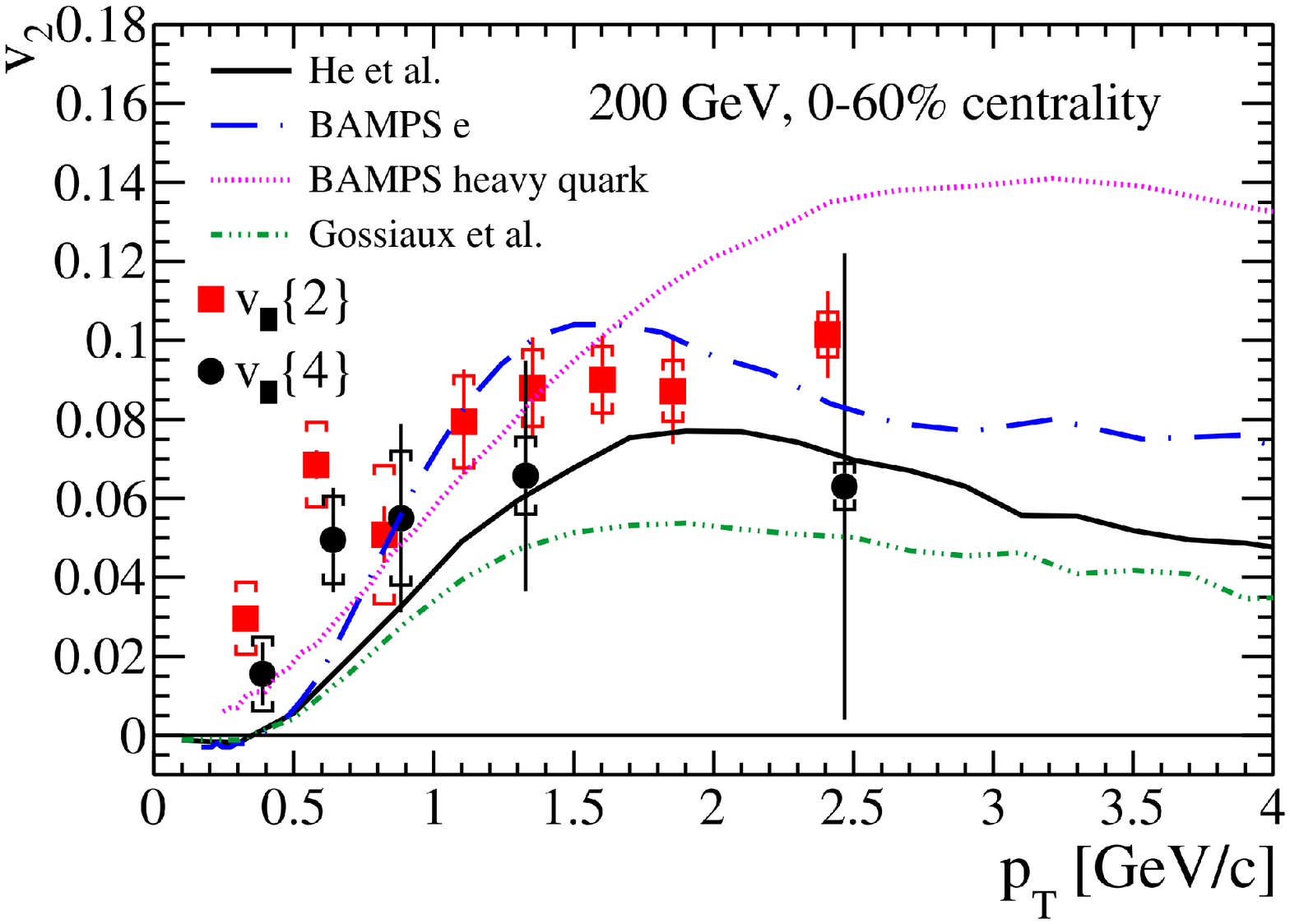}
\end{center}
\caption{{\bf Left}: Suppression of the heavy-flavour production at forward rapidity in d+Au collisions at PHENIX. 
{\bf Right}: Non-photonic electron azimuthal anisotropies in Au+Au collisions at STAR.}
\label{fig1}
\end{figure}

\begin{figure}[ht]
\begin{center}
\includegraphics[width=6.5cm]{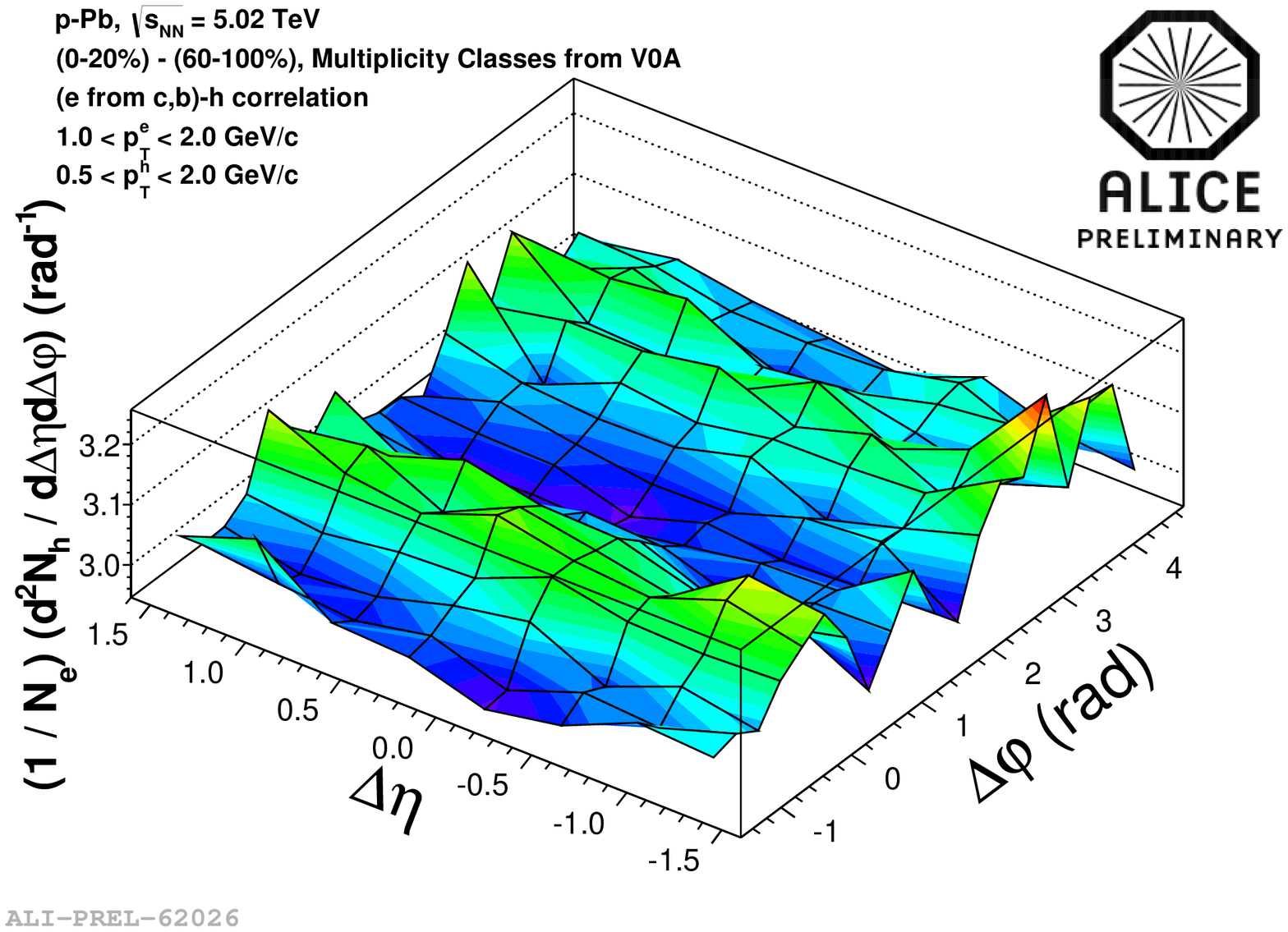}
\includegraphics[width=4.5cm]{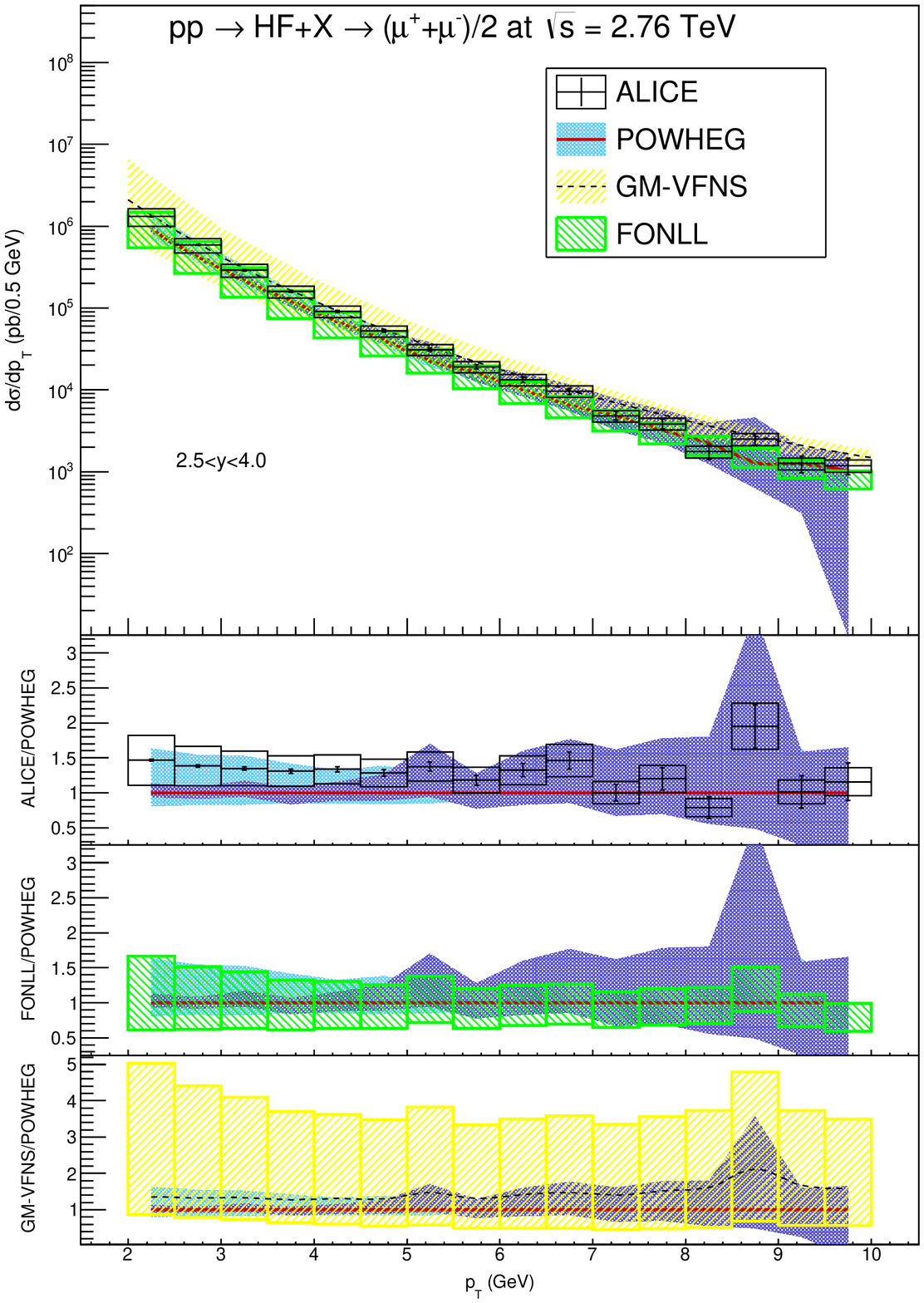}
\end{center}
\caption{{\bf Left}: Two-particle correlation distribution in p+Pb collisions at ALICE. {\bf Right}: Transverse-momentum distributions of muons from heavy-flavour 
(charm and bottom-quark) decay produced in the forward region at the LHC with $\sqrt{s}$= 7 TeV and compared to ALICE data.}
\label{fig2}
\end{figure}

Heavy flavors are suggested as excellent probes to study the properties 
of the hot and dense nuclear matter created in high-energy heavy ion collisions 
in the Relativistic Heavy Ion Collider (RHIC) and ALICE experiments.
These measurements will be a key to understand evolution 
of medium effects from proton-proton ($pp$) to heavy-ion collisions.
S.~Lim presented measurements~\cite{Adare:2013lkk,Adare:2014keg} 
of the transverse momentum $p_T$ of leptons decaying 
from charmed/beauty hadrons in deuteron + gold $(d+Au)$ 
collisions at the PHENIX experiment, and discussed several model calculations.
Results in Fig.\ref{fig1}(left) show suppression of the heavy-flavour pair production at forward rapidity.
Z.~Ye presented recent results~\cite{Adamczyk:2014yew} on open heavy-flavor production 
through semi-leptonic decay channels from the STAR experiment.
The results show a strong suppression for the Non-Photonic 
Electron (NPE) production at $\sqrt{s_{NN}}=200$ GeV in gold-gold $(Au+Au)$ collisions.
In Fig.\ref{fig1}(right) Non-photonic electron azimuthal anisotropies are compared 
to different model calculations at $\sqrt{s_{NN}}=200$ GeV.
D.~Thomas presented measurements~\cite{Filho:2014vba} of open heavy-flavour production 
cross sections and their dependence on charged particle multiplicity 
in $pp$ ($\sqrt{s}=2.76$ TeV and $\sqrt{s}=7$ TeV) and proton-lead ($p+Pb$) collisions 
($\sqrt{s}=5.02$ TeV) at the ALICE experiment. 
Differences between two-particle correlation distribution in high (0-20\%) 
and low (60-100\%) multiplicity in the $(\Delta\eta,\Delta\phi)$ space, are shown in Fig.\ref{fig2}(left), where
a double-ridge structure is observed in heavy-flavour decay electron azimuthal 
angular correlations in $p+Pb$ collisions. 
K.~Kovarik presented a theory talk in which theoretical predictions obtained with shower Monte Carlo programs 
are compared to different theories where fixed-order calculations are extended with next-to-leading logarithms 
and to predictions using the General Mass Variable Flavour Number Scheme (GM-VFNS). 
Results~\cite{Klasen:2014dba} are compared to recent measurements of heavy-flavour production in $pp$ collisions 
at the ALICE experiment~\cite{Abelev:2012pi} in Fig.\ref{fig2}(right). 
These findings are important to understand theory uncertainty 
for the baseline processes before studying the heavy-quark suppression in heavy-ion collisions.

\section{B-physics and charm/beauty rare decays}

\begin{figure}[ht]
\begin{center}
\includegraphics[width=8cm]{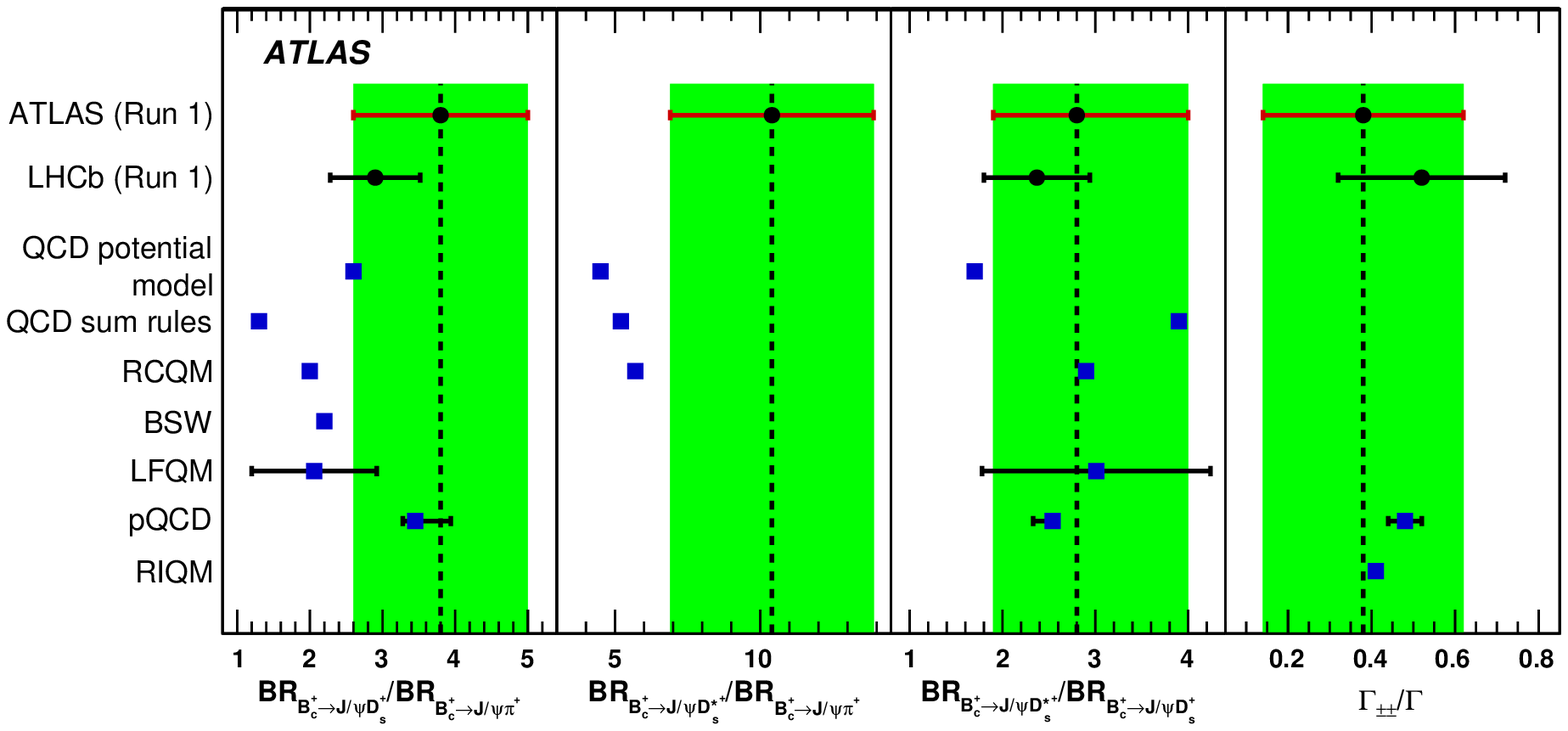}
\includegraphics[width=6cm]{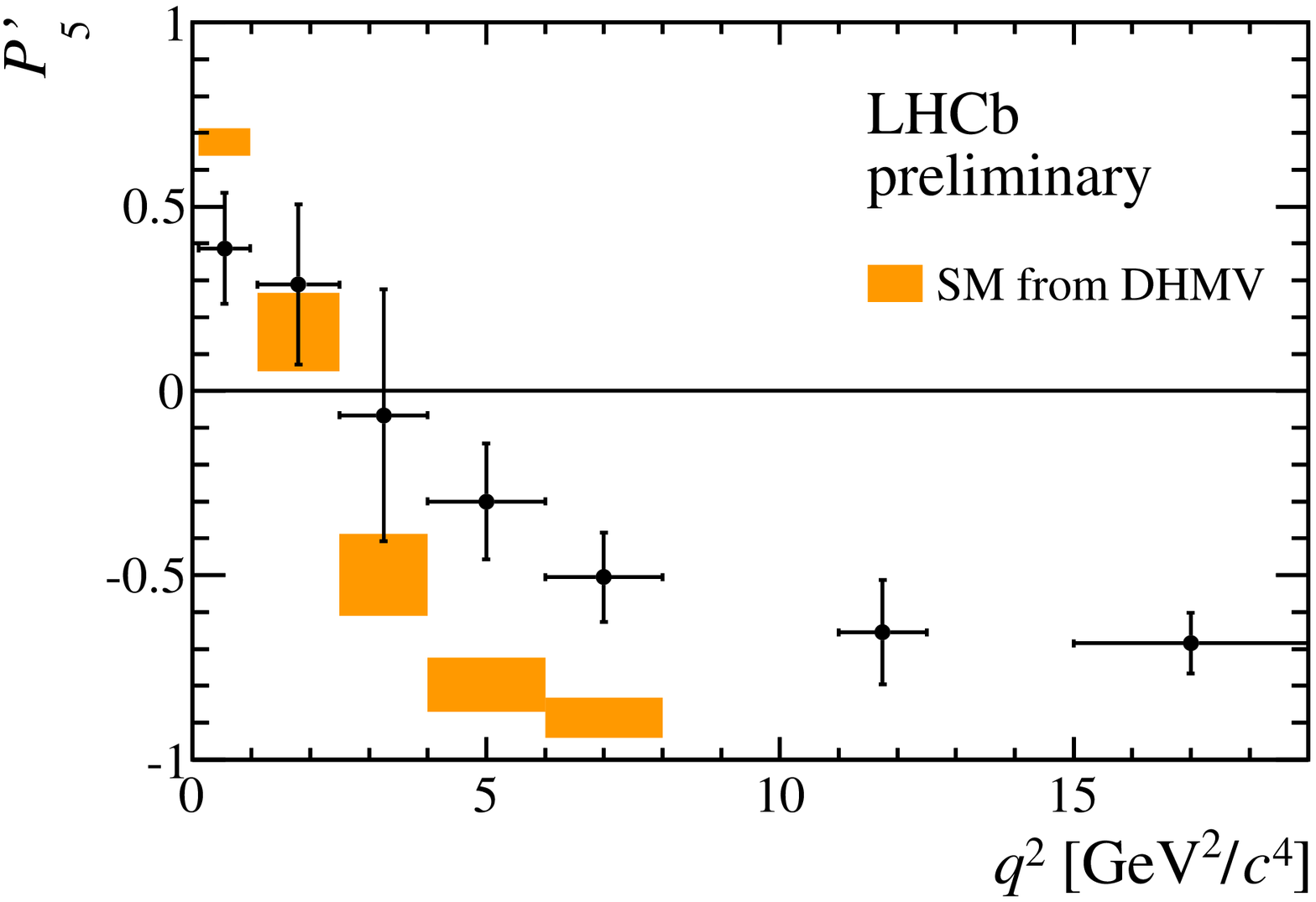}
\end{center}
\caption{{\bf Left}: Comparison of the results of this measurement with those of LHCb and theoretical predictions based on a QCD relativistic potential model. 
{\bf Right}: The observable ${P'}_5$ in bins of $q^2$. The shaded boxes show the SM prediction of Ref.~\cite{Descotes-Genon:2014uoa}}
\label{fig3}
\end{figure}

\begin{figure}[ht]
\begin{center}
\includegraphics[width=5cm]{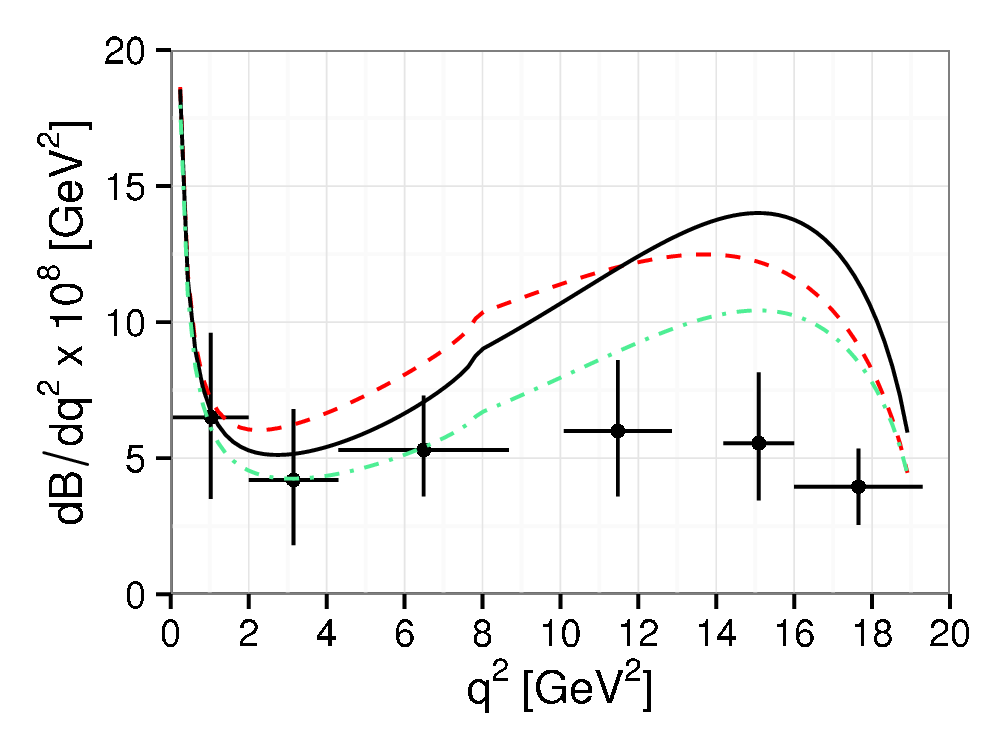}
\includegraphics[width=4.5cm]{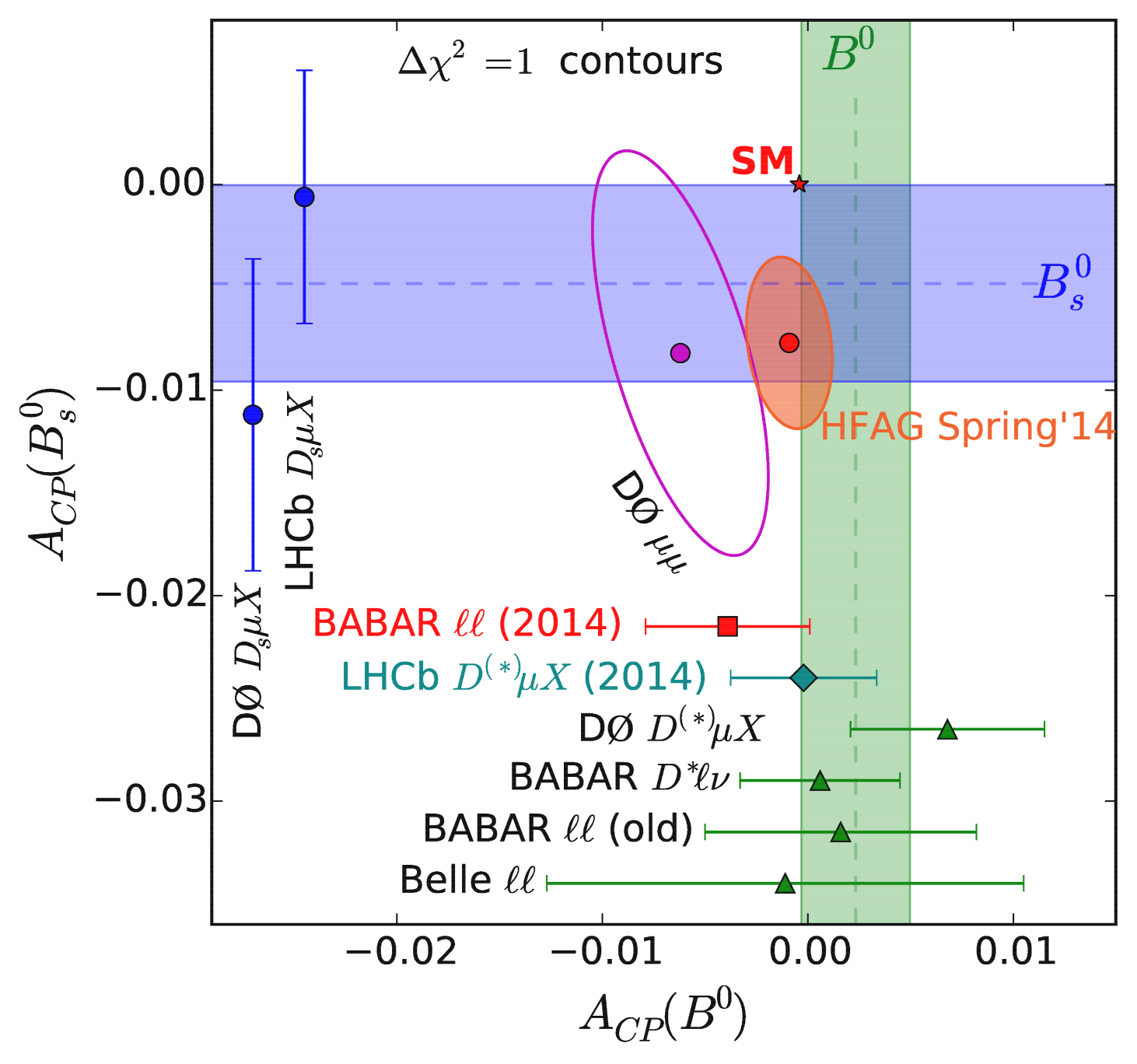}
\includegraphics[width=4.5cm]{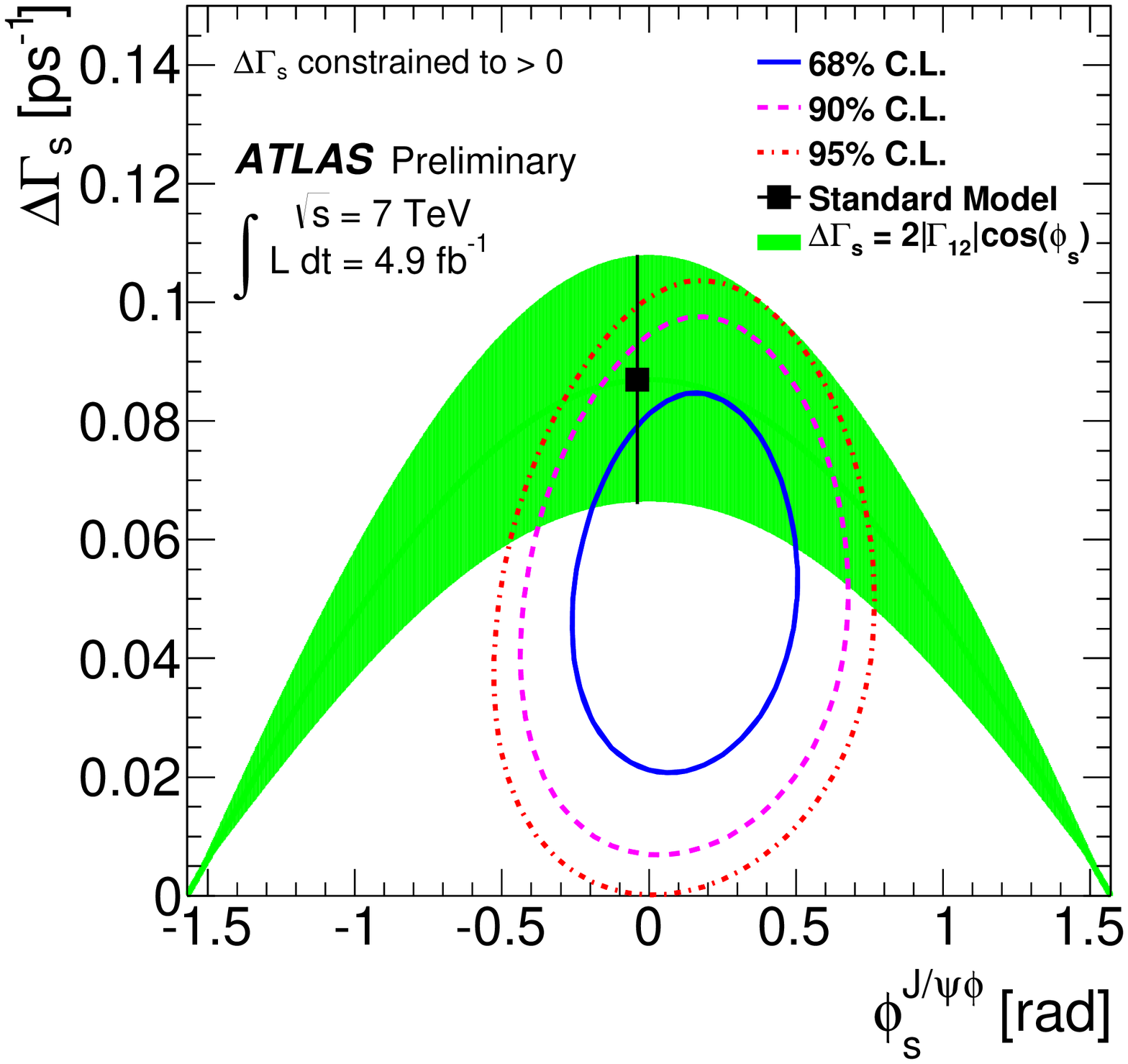}
\end{center}
\caption{{\bf Left}: The AdS/QCD prediction for $B \rightarrow K^*\mu^+ \mu^-$ differential decay width.
{\bf Center}: Measurements of CP asymmetry in neutral B mixing, including this measurement, recent LHCb result.
{\bf Right}: Likelihood contours in $\phi_s - \Delta\Gamma_s$ plane. The blue and red contours show the 68\% and 95\% likelihood contours, 
respectively (statistical errors only). The green band is the theoretical prediction of mixing-induced CP violation.}
\label{fig4}
\end{figure}

Several presentations focused on the physics of the B-meson 
and rare decays of charm and bottom-quark decays. 
B-physics and heavy-flavour rare decays are extremely important 
as they lead to interplays of the flavour sector and collider physics.
Moreover, precise measurements in the flavour sector set severe constraints on new physics.
New measurements~\cite{Aad:2015eza} of B-hadron production with the ATLAS experiment were presented by J.~Zalieckas. He showed 
studies of $B^{+}_c\rightarrow J/\psi D^{(*)+}_s$ using $\sqrt{s}$ = 7 and 8 TeV data collected in 2011 and 2012.
The main results are shown in Fig.\ref{fig3}(left) where branching ratios seem to be 
generally well described by perturbative QCD (pQCD) and are consistent with several other theoretical models such as QCD potential~\cite{Colangelo:1999zn}, and Light-front quark model (LFQM)~\cite{Ke:2013yka}.
These results are found to be consistent with similar measurements~\cite{Aaij:2013gia} performed at the LHCb experiment.
M.~Chrzaszcz presented rare charm and bottom decays measurements~\cite{LHCb:2015dla} using $pp$ collision data at the LHCb experiment.
In particular, he discussed the angular analysis of the $B^{0} \rightarrow K^{*0} \mu^{+} \mu^{-}$ decays and 
results for a relevant angular variable are shown in Fig.\ref{fig3}(right).
Neglecting the correlations between the observables, the measurements seem to be in agreement with the SM predictions, 
although the ${P'}_5$ angular observable exhibits a local tension with respect to the SM prediction at a level of 3.7$\sigma$.
These results have important implications on the limits on anomalous triple gauge bosons couplings.
New theoretical models for B-decays were discussed in M.~Ahmady's presentaion~\cite{Ahmady:2014cpa}. 
He discussed results obtained from anti-de Sitter Quantum Chromodynamics (AdS/QCD) and 
used to calculate light cone distribution amplitudes for $\rho$ and $K^*$ vector mesons.
In Fig.\ref{fig4}(left) these distribution amplitudes, utilized to calculate the $B \rightarrow K^*\mu^+ \mu^-$ 
differential decay width, are compared to recent LHCb data.
CP asymmetries between same-sign inclusive dilepton samples $l^+ l^+$ and $l^- l^-$ $(l = e,\mu)$ from semileptonic 
B decays in $\Upsilon(4S) \rightarrow  B\bar{B}$ events have been measured~\cite{Lees:2014kep} 
at the BaBar experiment and presented by M.~Chrzaszcz.
These asymmetries are important probes for CP and T symmetry violations. 
Results are shown in Fig.\ref{fig4}(center) where the CP asymmetry is $A_{CP} = (-3.9 \pm 3.5(stat.) \pm 1.9(syst.)) \times 10^{-3}$ and is 
consistent with the SM expectation.
T.~Nooney presented results of new physics searches with B mesons at the ATLAS experiment~\cite{Aad:2014cqa}. In particular, 
she discussed the parameters of $B_s \rightarrow J/\psi\phi$ decay and showed that they are consistent with the values predicted by the SM.
Fig.\ref{fig4}(right) shows the main results of this study, where the width difference $\Delta \Gamma_s$ 
of the two $B^0_s$ and $\bar{B}^0_s$ mesons is represented in terms of the CP violating phase $\phi_s$. 
Many new physics models can significantly affect $\phi_s$, therefore this observable is suitable to search for BSM phenomena.

\section{Charm and beauty quark production and exotic mesons}

\begin{figure}[ht]
\begin{center}
\includegraphics[width=5cm]{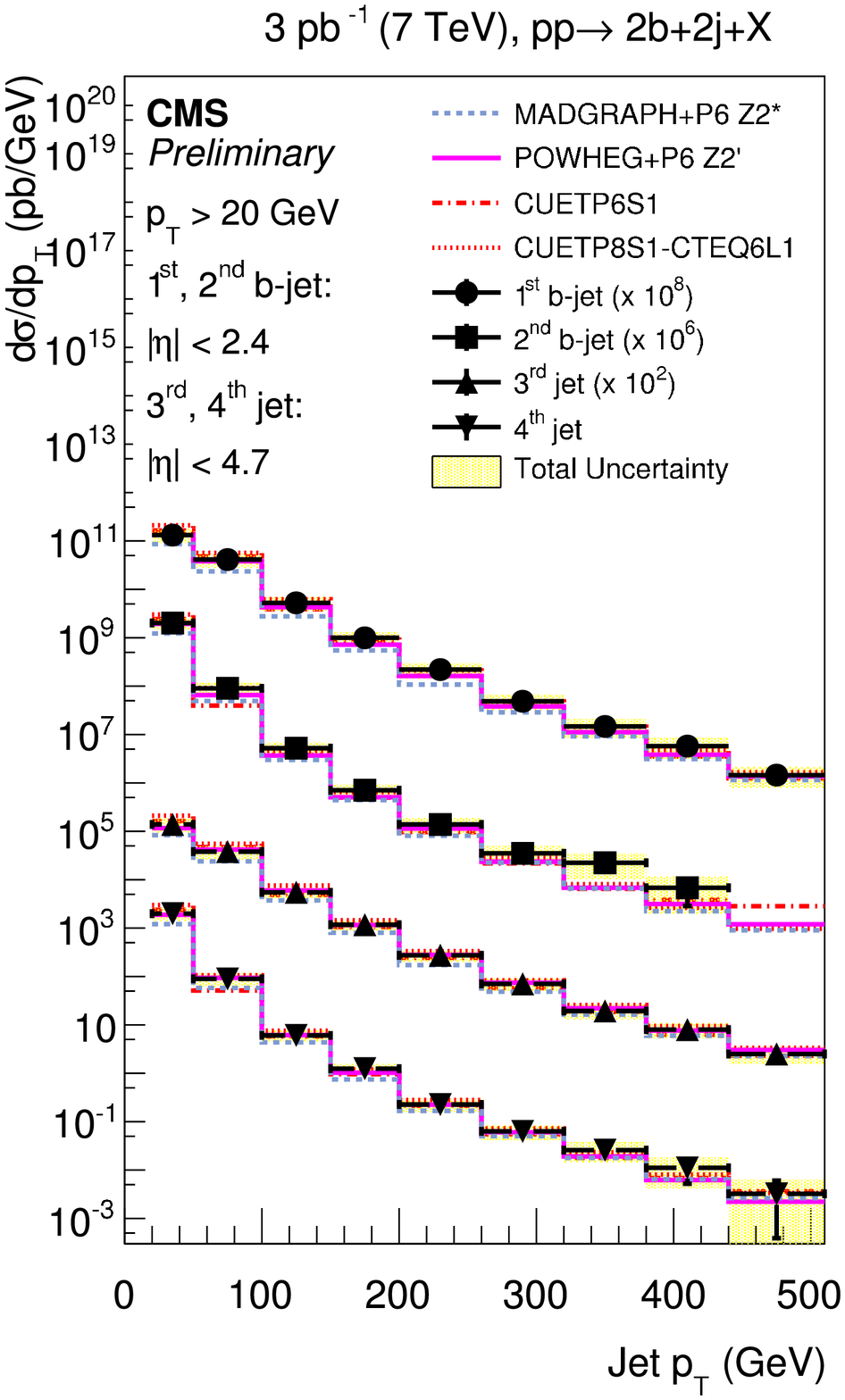}
\includegraphics[width=7cm]{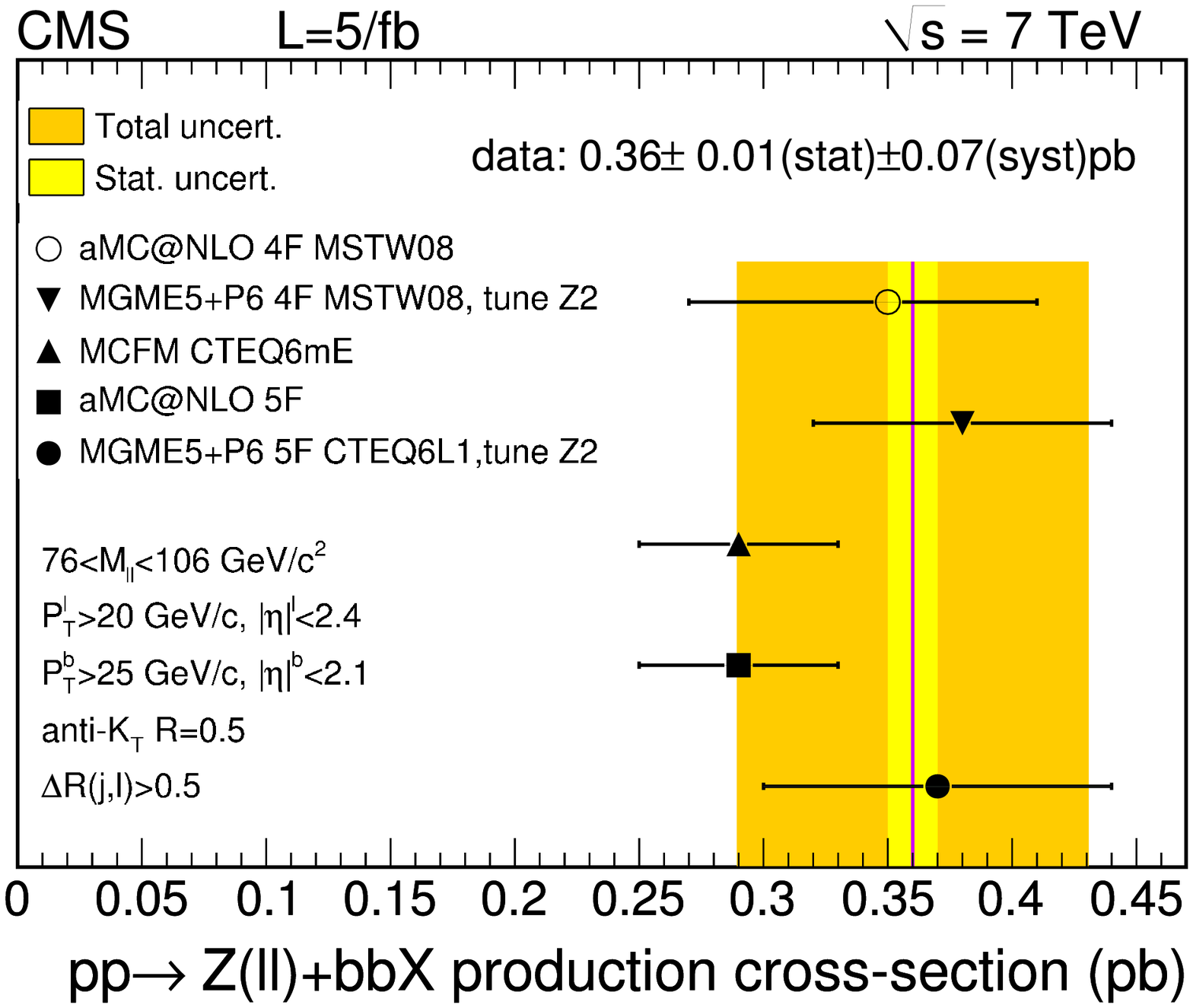}
\end{center}
\caption{{\bf Left}: Differential cross sections unfolded to the stable particle level as a function of the jet transverse momenta at CMS 7 TeV.
{\bf Right}: Total inclusive cross section for $pp\rightarrow Z(l\bar{l}) + b\bar{b} X$ at CMS 7 TeV.}
\label{fig5}
\end{figure}

\begin{figure}[ht]
\begin{center}
\includegraphics[width=5cm]{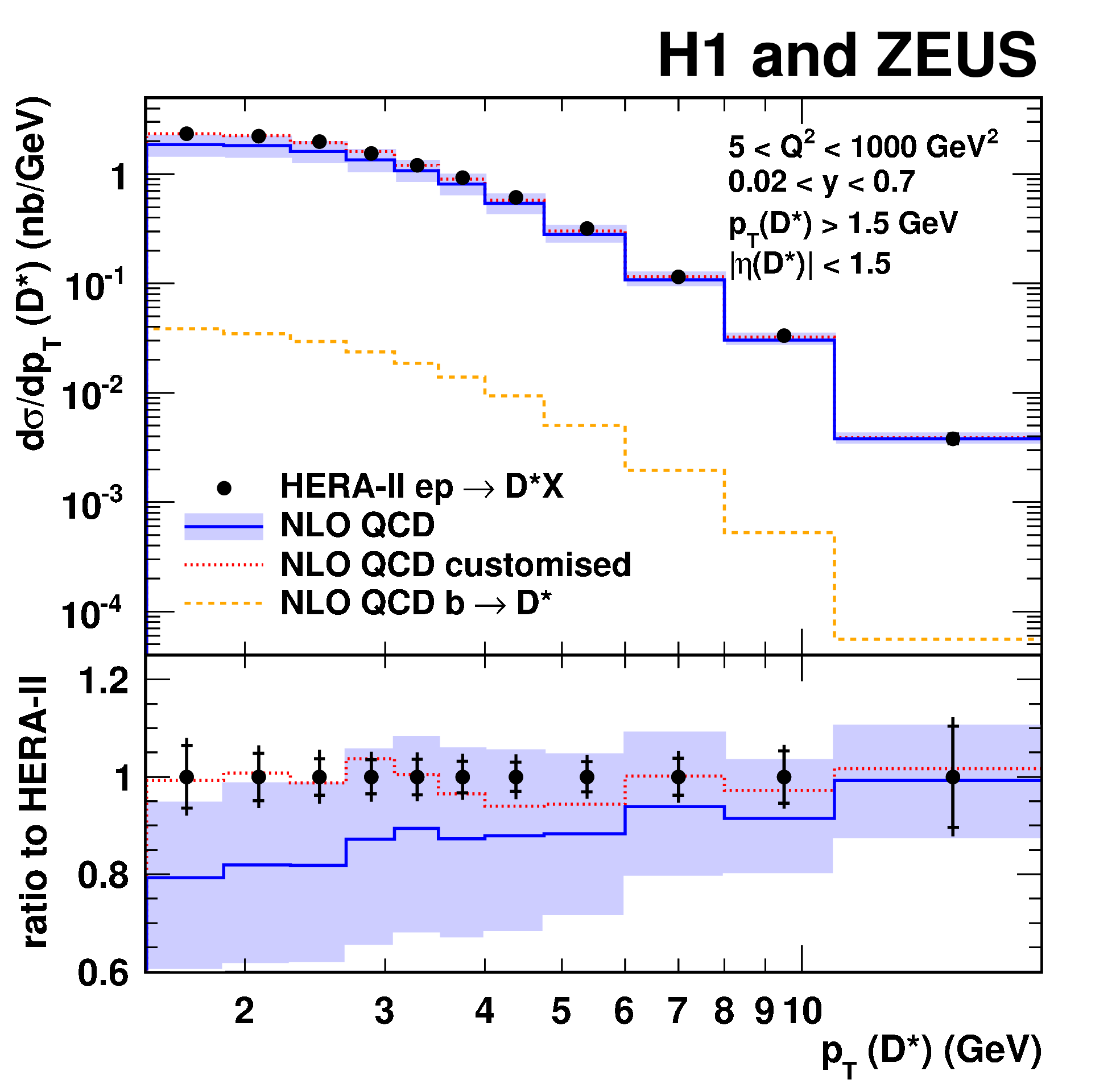}
\includegraphics[width=6cm]{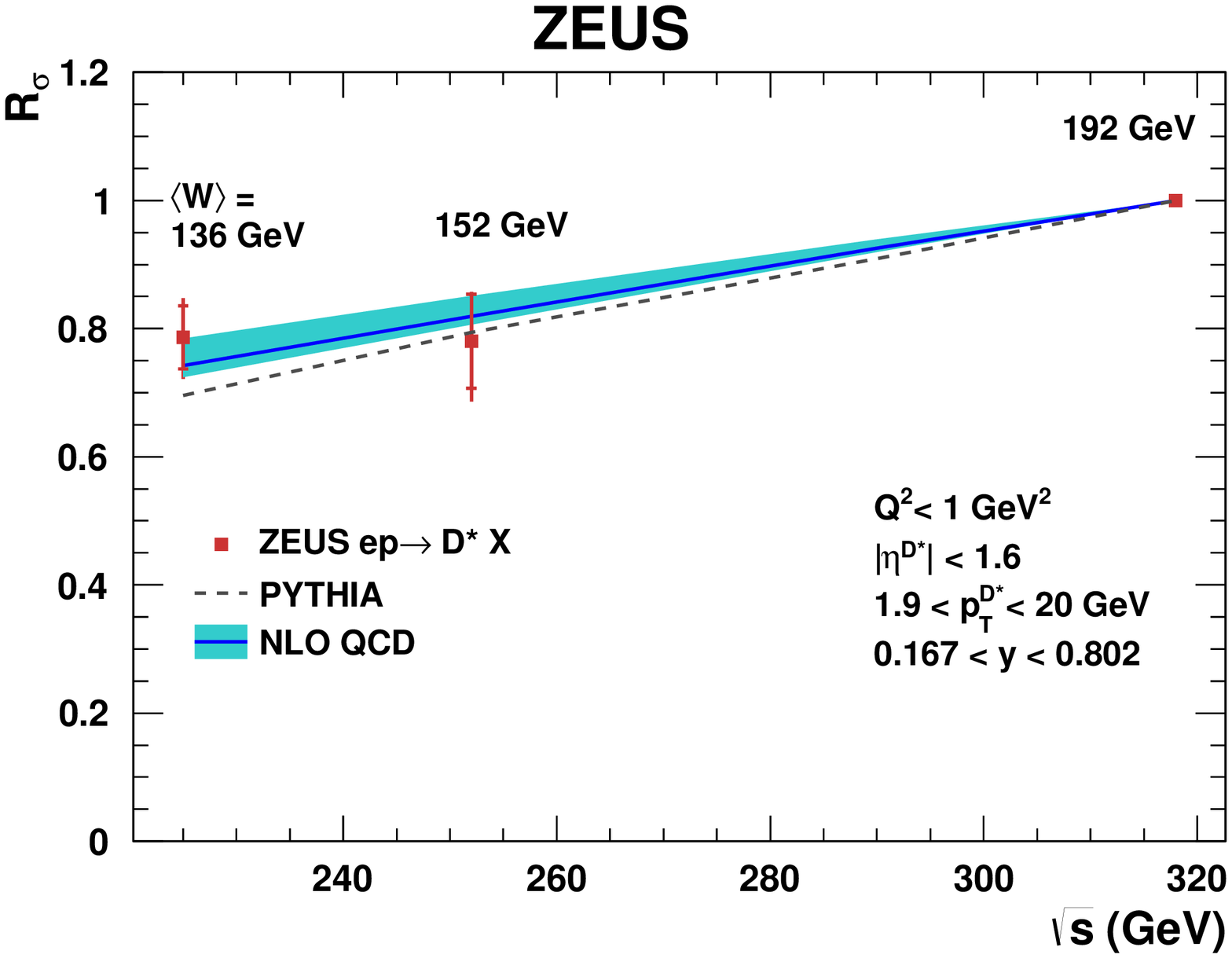}
\end{center}
\caption{{\bf Left}: Differential $D^{(*)\pm}$-production cross section as a function of $p_T (D^∗)$ at HERA. The data points
are the combined cross sections. 
{\bf Right}: Normalised $D^{(*)}$ visible photoproduction cross sections as a function of the $ep$ centre-of-mass energy at HERA.}
\label{fig6}
\end{figure}

\begin{figure}[ht]
\begin{center}
\includegraphics[width=5cm]{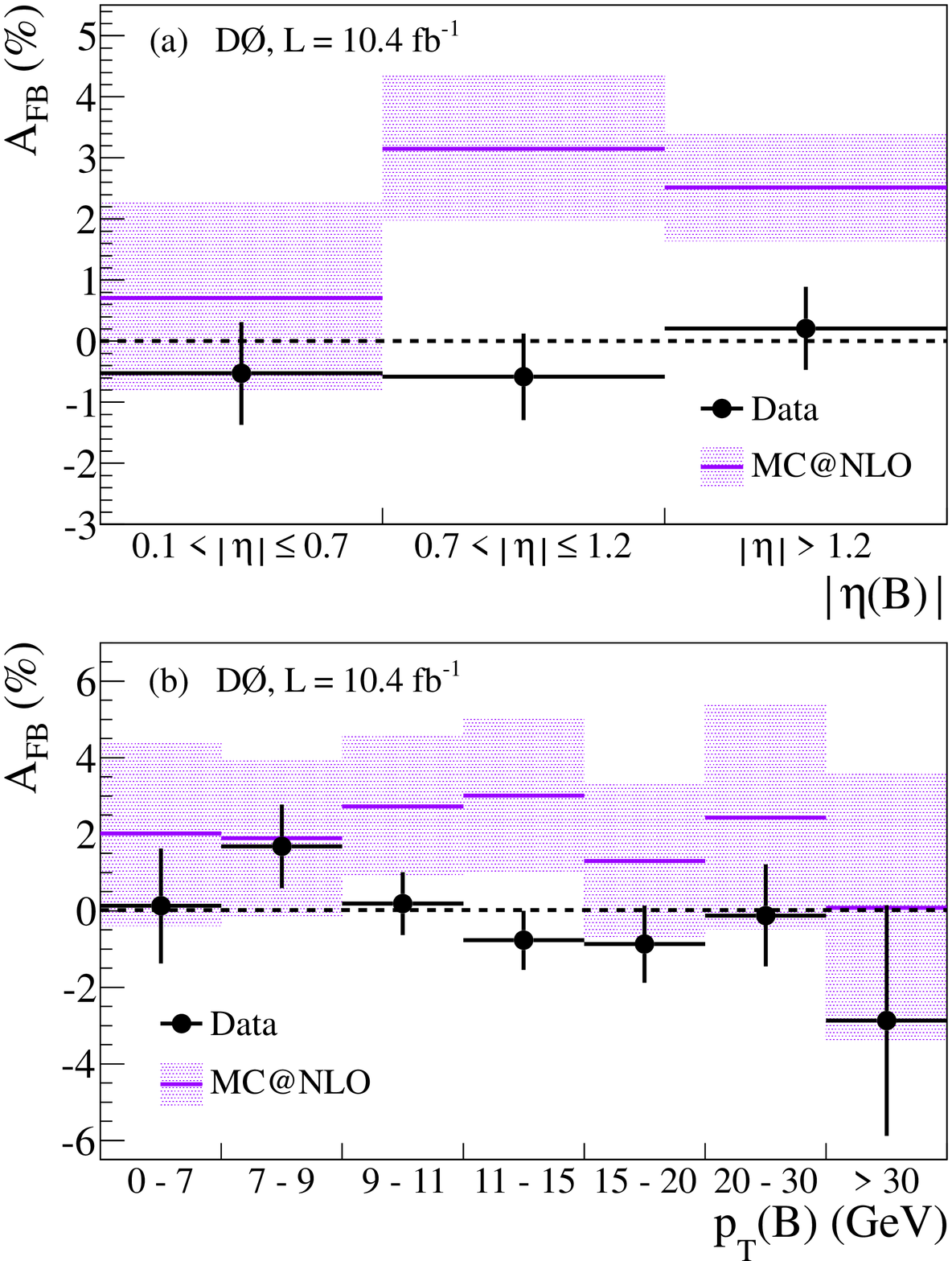}
\includegraphics[width=6cm]{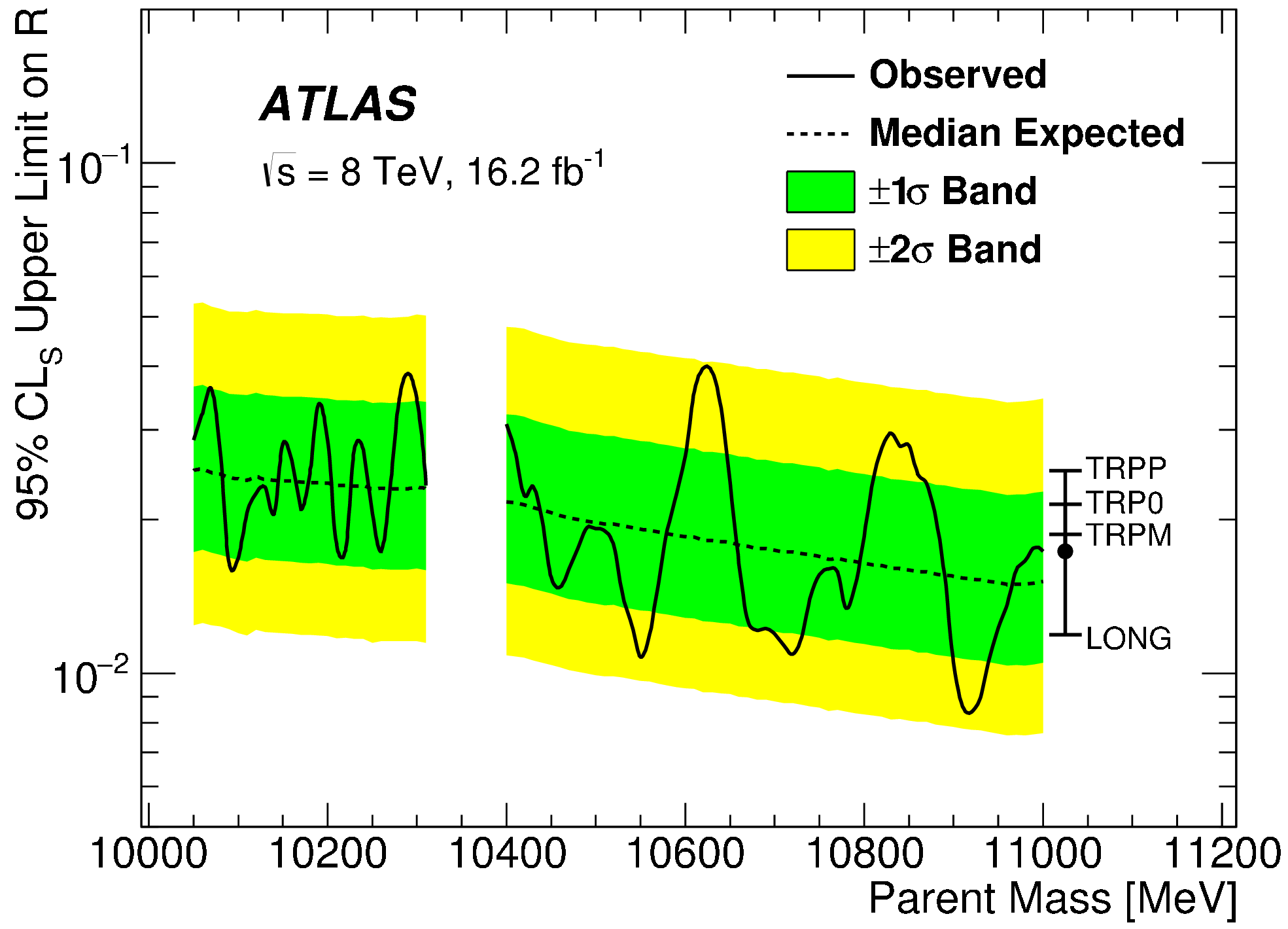}
\end{center}
\caption{{\bf Left}: Comparison of $A_{FB}(B^{\pm})$ and $A^{SM}_{FB}(B^{\pm})$ in bins of $|\eta_B|$ and $p_T(B)$. 
Data points and MC bands include statistical uncertainties convoluted with systematic uncertainties.
{\bf Right}: Observed 95\% CL upper limits (solid line) on the relative production 
rate of a hypothetical $X_b$ parent state decaying isotropically to $\pi^+$ $\pi^-$ $\Upsilon$, as a function of mass.}
\label{fig7}
\end{figure}

The charm and beauty-quark production process is going to become extremely important for future precision programs at colliders.
Differential cross sections for the production of charm and beauty-quark in QCD jets, or in association with vector bosons,
are excellent probes of QCD factorization and also have the potential to constrain PDFs of the proton, if such measurements are sufficiently precise.
The ATLAS, CMS, Tevatron, and HERA experiments presented several new results for these observables.
P.~Gunnellini presented new results for the measurement of four-jet production including 
two b-quark jets at the CMS experiment~\cite{CMS:2015dna} at $\sqrt{s}=$ 7 TeV. 
The main results of this analysis are shown in Fig.\ref{fig5}(left), where the differential 
cross sections as a function of the $p_T$ of the jet is compared to the SM calculation in different rapidity bins.    
The SM prediction is found to be in good agreement with the data.
P.T.~Mastrianni showed recent measurements beauty-quark pair production associated with a vector boson at the CMS~\cite{Chatrchyan:2014dha,Chatrchyan:2013uza,Chatrchyan:2013uja} experiment at $\sqrt{s}$= 7 TeV.
In particular he discussed the total inclusive cross section for $pp\rightarrow Z(l\bar{l}) + b\bar{b} X$ process, illustrated in Fig.\ref{fig5}(right), 
and compared the experimental measurements to several theoretical QCD calculations at the NLO accuracy. 
Data and theory are in very good agreement within the uncertainties.  
Measurements of open charm production in deep-inelastic electron-proton ($ep$) scattering (DIS) at HERA 
provide important input for stringent tests of the theory of strong interactions.
Moreover these precise measurements provide a consistent determination of several important physical quantities such as
the charm contribution to the proton structure functions, the 
charm-quark mass $m_c$, and allow us to obtain improved predictions for $W$ and $Z$-production cross sections at the LHC. 
Several new results from HERA have been presented in the current workshop. 
A combination of differential $D^{(*)\pm}$ cross-section measurements from H1 and ZEUS collaborations at HERA~\cite{H1:2015dma} 
was presented by O.~Behnke. Perturbative next-to-leading-order QCD predictions are compared to the results for the $p_T$ spectrum of the $D^*$ in Fig.\ref{fig6}(left). 
The predictions describe the data well within their uncertainties, although higher order calculations 
would be helpful to reduce the theory uncertainty to a level more comparable with the data precision. Further improvements in the
treatment of heavy-quark fragmentation would also be desirable.
M.~Wing presented measurements of $D^{(*)}$ photoproduction at three different centre-of-mass energies from the ZEUS collaboration at HERA~\cite{Abramowicz:2014ihk}.
The dependence on the $ep$ centre-of-mass energy was presented for the first time. 
Variations of the cross section with centre-of-mass energy are sensitive to the gluon PDF in the proton, 
as different values of Bjorken $x$ are probed. The main results are shown in Fig.\ref{fig6}(right) where the 
normalised $D^{(*)}$ visible photoproduction cross sections as a function of the $ep$ centre-of-mass energy are compared to NLO QCD theory 
predictions. Measurements are in good agreement with QCD predictions. 
B.K.~Abbott presented for the first time a new measurement of the forward-backward 
asymmetry ($A_{FB}$) in the production of $B^{\pm}$ mesons in $p\bar{p}$ collisions 
at the D0 experiment during the Run-II of the Tevatron collider~\cite{Abazov:2015vea,Abazov:2014ysa}. 
He showed that D0 measured no significant forward-backward asymmetry. The result for 
$A_{FB}(B^{\pm})=-0.0024\pm 0.0041 (stat)\pm 0.0019 (sys)$ illustrated in Fig.~\ref{fig7}(left), 
is compatible with zero. Locally, some tensions between data and the MC@NLO~\cite{Frixione:2002ik} theory prediction are found.
Nonzero asymmetries would indicate a preference for a particular flavor, i.e., b quark or b antiquark, to be produced 
in the direction of the proton beam. These measurements provide important constraints on the production mechanisms of heavy quarks at hadron colliders.

Searches of exotic states are going on at the LHC. In particular the existence of the $X(3872)$ resonance suggests 
the presence of its bottomonium counterpart $X_b$. J.M.~Izen presented a study in which 
searches for $X_b$ states with the ATLAS experiment~\cite{Aad:2014ama} in final states including $\pi^+$ $\pi^-$ $\Upsilon$ channel, are discussed.
The main results are shown in Fig.\ref{fig7}(right), in which exclusion limits for the relative production rate of the $X_b$ state are given as a function of its mass. 

\section{Top-quark physics: production and properties}

\begin{figure}[ht]
\begin{center}
\includegraphics[width=6cm]{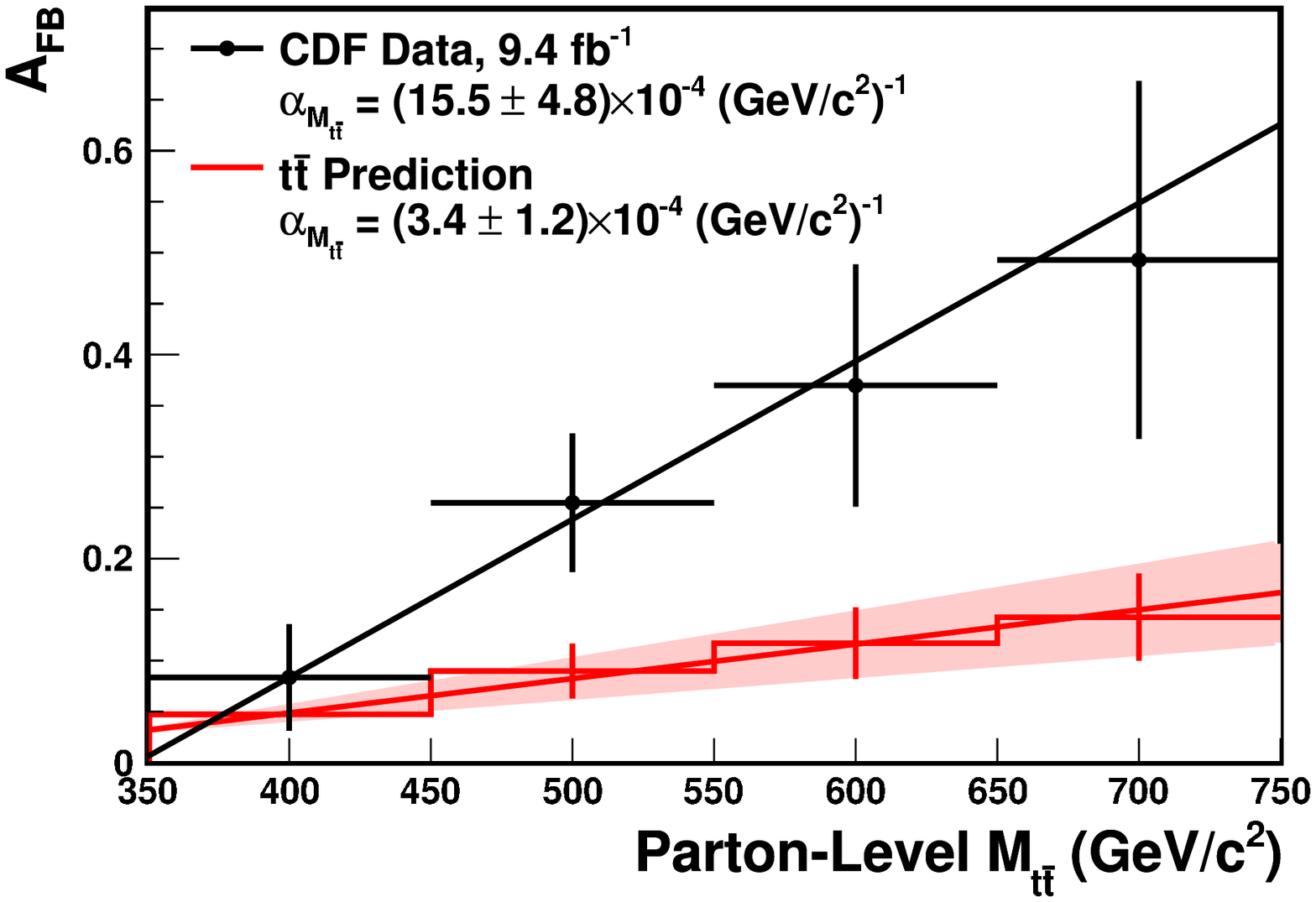}
\includegraphics[width=6cm]{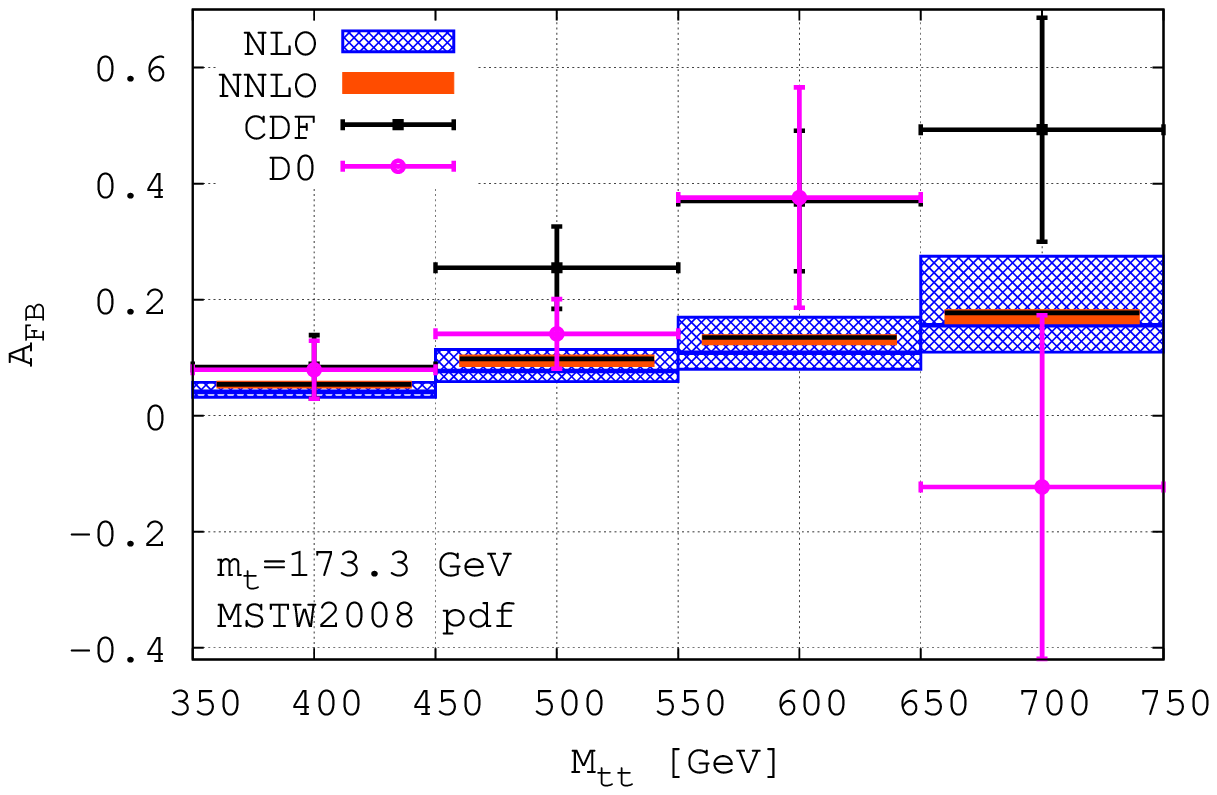}
\end{center}
\caption{{\bf Left}: The parton-level $M_{t\bar{t}}$ distribution for the $A_{FB}$ at CDF~\cite{Aaltonen:2012it}. 
The shaded region represents the theoretical uncertainty on the slope of the prediction.
{\bf Right}: $M_{t\bar{t}}$ distribution for the $A_{FB}$ in pure QCD at NLO (blue) 
and NNLO (orange) versus CDF~\cite{Aaltonen:2012it} and D0~\cite{Abazov:2014cca} data.}
\label{fig8}
\end{figure}

The top-quark physics sessions of the current workshop were very active and triggered several interesting discussions. 
New results were presented from both experimental collaborations and theorists.
J.~Wilson presented new results for the measurements of 
differential forward-backward asymmetries as a function of rapidity $\Delta y = y_t - y_{\bar{t}}$ 
and invariant mass of bottom- and top-quark pair production at CDF Tevatron~\cite{Aaltonen:2015mba}.
Recent measurements for the $M_{t\bar{t}}$ distribution for $A_{FB}$ are illustrated in Fig.\ref{fig8}(left) where these are compared to the POWHEG~\cite{Frixione:2007nw} theory prediction.
It has been recently found~\cite{Czakon:2014xsa} that the NNLO QCD corrections to the top-quark pair production cross section 
are large (27\% of the NLO) and when these are combined with the EW (25\% of the NLO) corrections, produce inclusive $A_{FB}\approx$ 10\%. 
The plot in Fig.\ref{fig8}(right), presented by A.~Mitov in the plenary session, 
shows the NNLO QCD prediction for the $M_{t\bar{t}}$ differential asymmetry together with the 
CDF~\cite{Aaltonen:2012it} and D0~\cite{Abazov:2014cca} measurements.    
The agreement between the data and theory has improved, and the NNLO theory is less than 1.5$\sigma$ below the CDF data.

\begin{figure}[ht]
\begin{center}
\includegraphics[width=6cm]{./Fig-Kidonakis.eps}
\includegraphics[width=7cm]{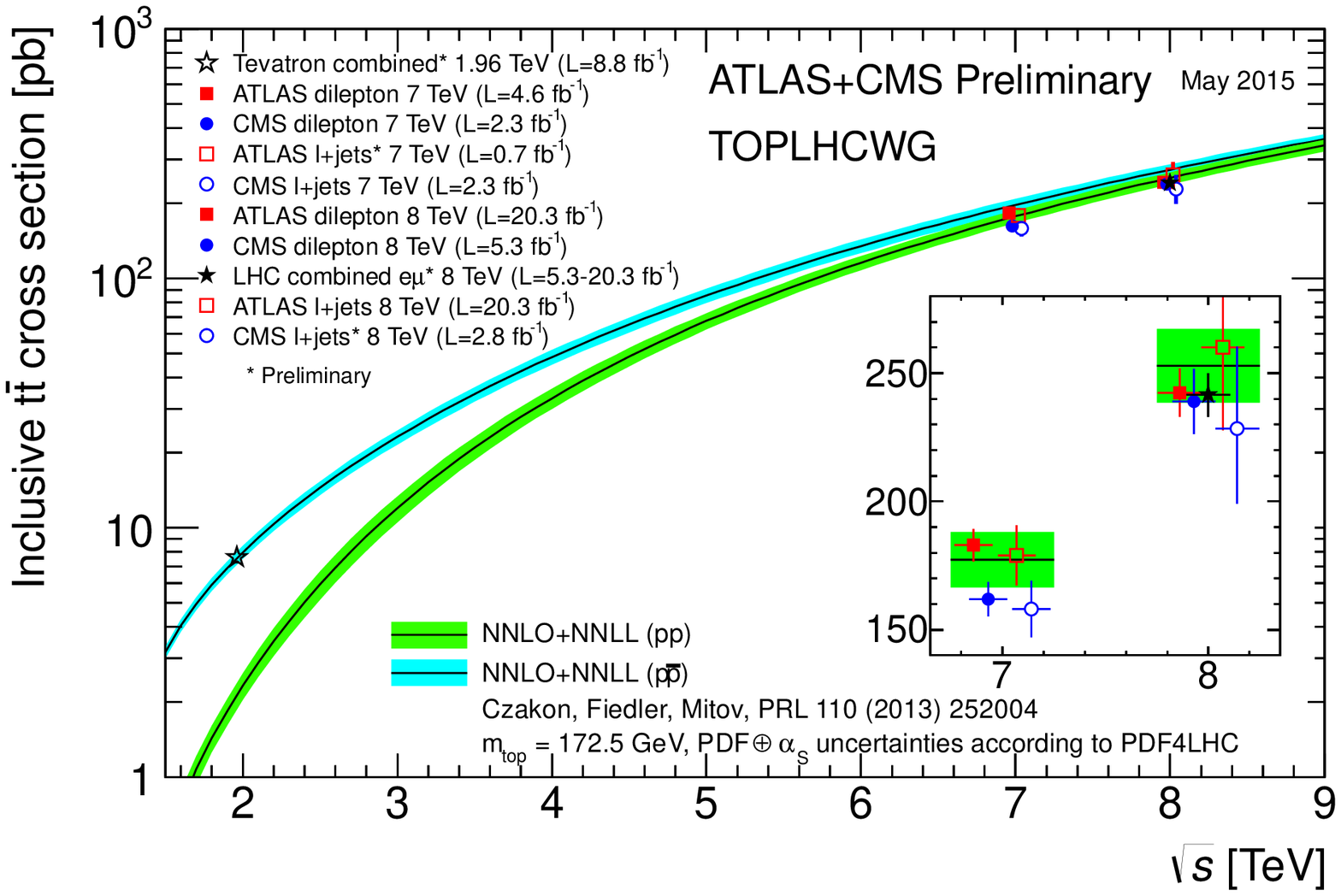}
\end{center}
\caption{{\bf Left}: 
Normalized aN3 LO top-quark pT distributions at the 7 TeV LHC, and comparison with CMS data~\cite{Chatrchyan:2012saa}.
{\bf Right}: Summary of LHC and Tevatron measurements of top-quark pair production.}
\label{fig9}
\end{figure}

N.~Kidonakis reported about the recent progress in approximate QCD calculations for top-quark pair 
production using logarithmic threshold expansions of the resummed cross section.
In particular, he showed approximate cross sections up to NNNLO (${\cal O}(\alpha_s^5)$) 
for relevant ${t\bar{t}}$ observables at the LHC and Tevatron~\cite{Kidonakis:2014pja}.
In Fig.\ref{fig9}(left) these higher-order corrections are shown and are compared to recent LHC measurements. 

\begin{figure}[ht]
\begin{center}
\includegraphics[width=6cm]{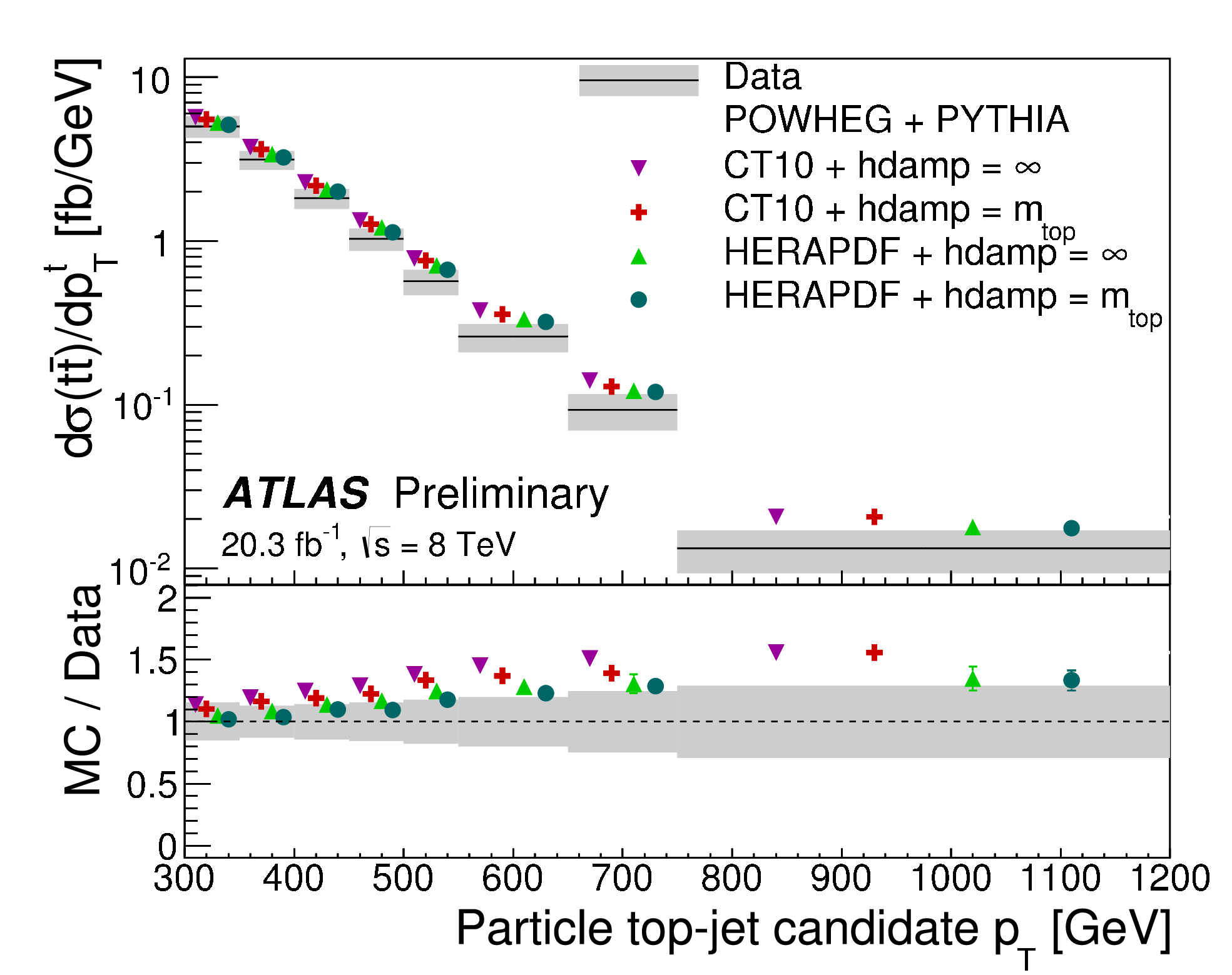}
\includegraphics[width=5.5cm]{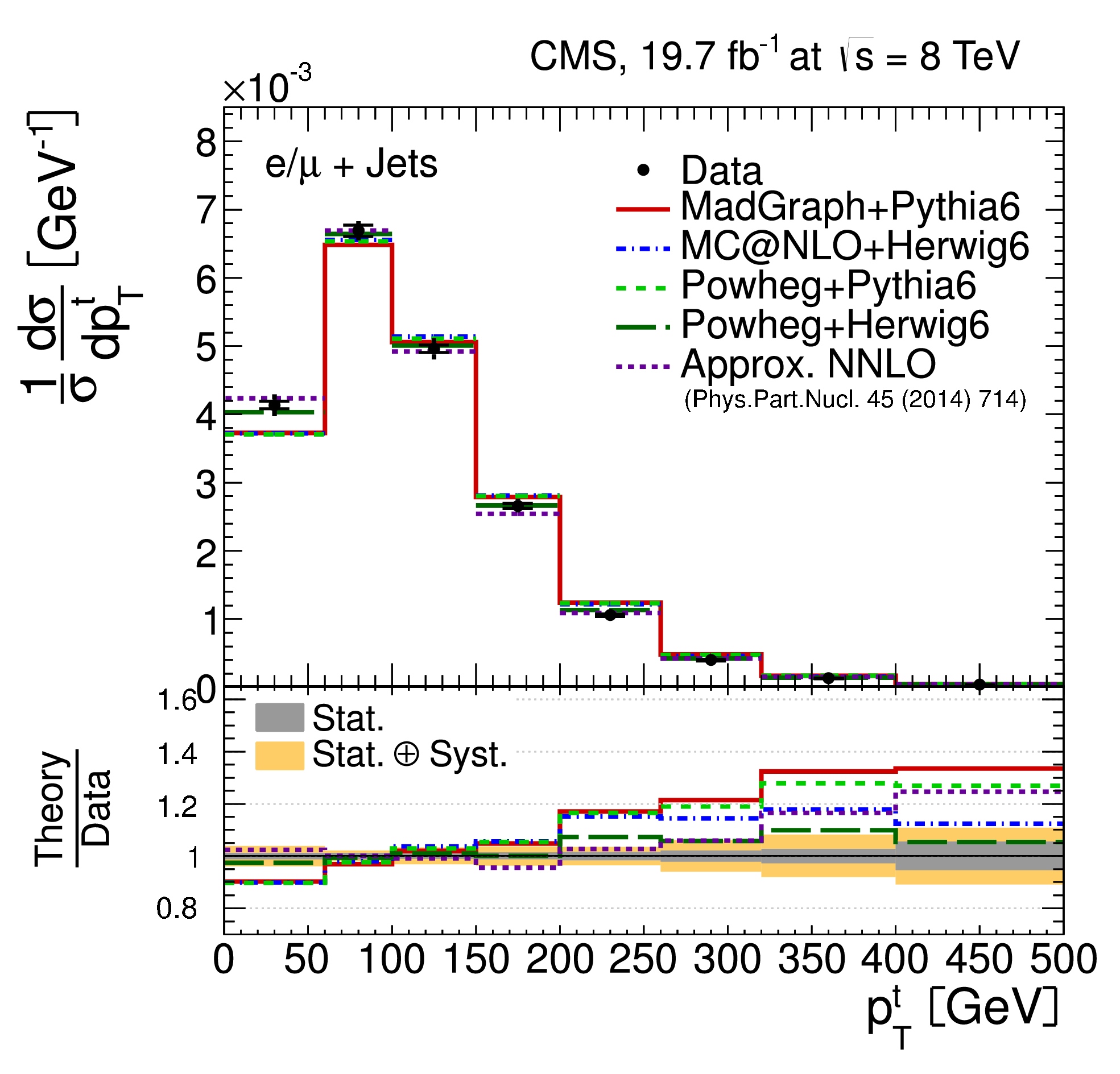}
\end{center}
\caption{{\bf Left}: Normalized differential cross-section of highly boosted top quarks as a function of the top quark $p_T$.    
{\bf Right}: Normalized differential ${t\bar{t}}$ production cross section in the $l$+jets channels as a function of the top-quark $p_T$
}
\label{fig10}
\end{figure}

\begin{figure}[ht]
\begin{center}
\includegraphics[width=7cm]{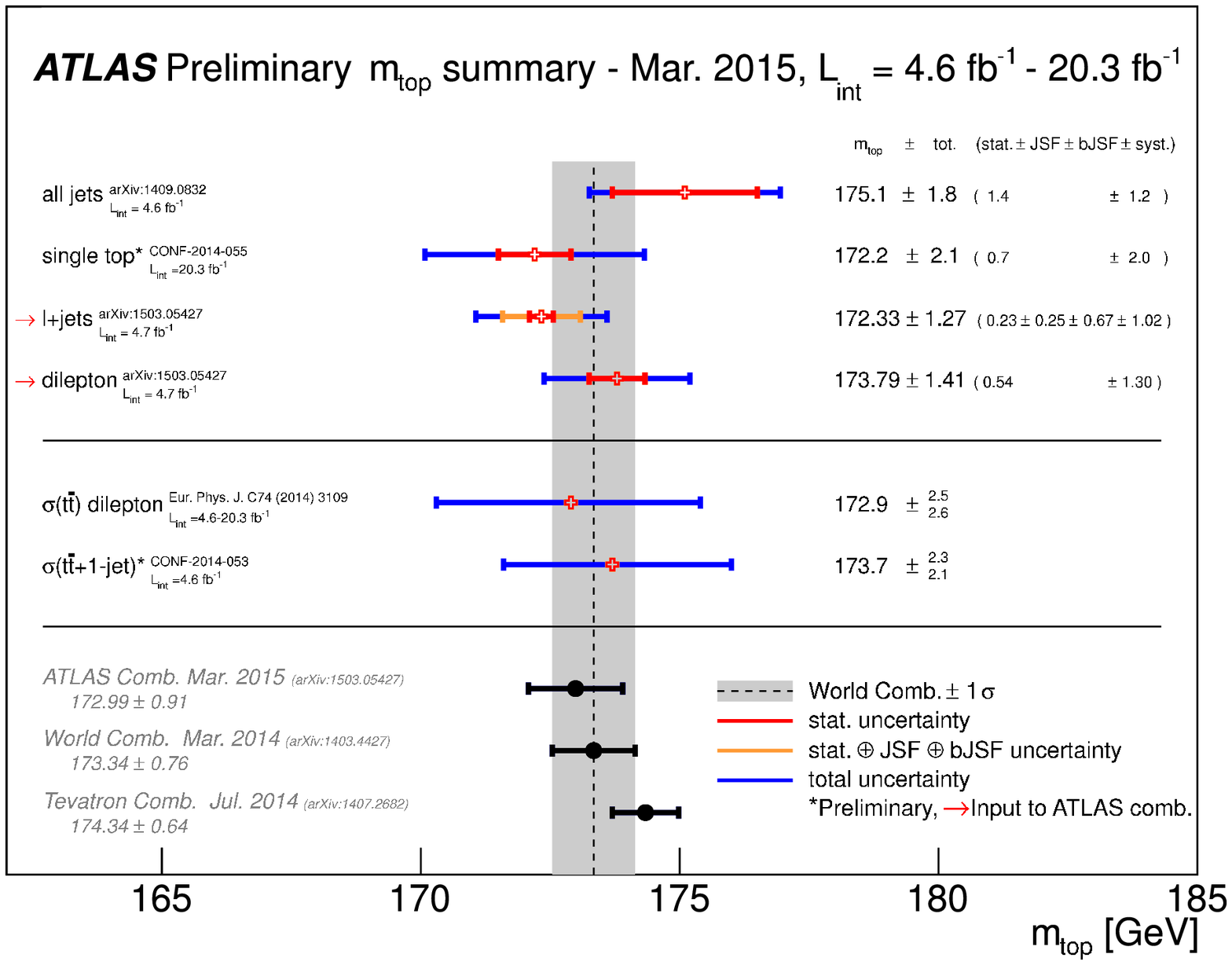}
\includegraphics[width=5cm]{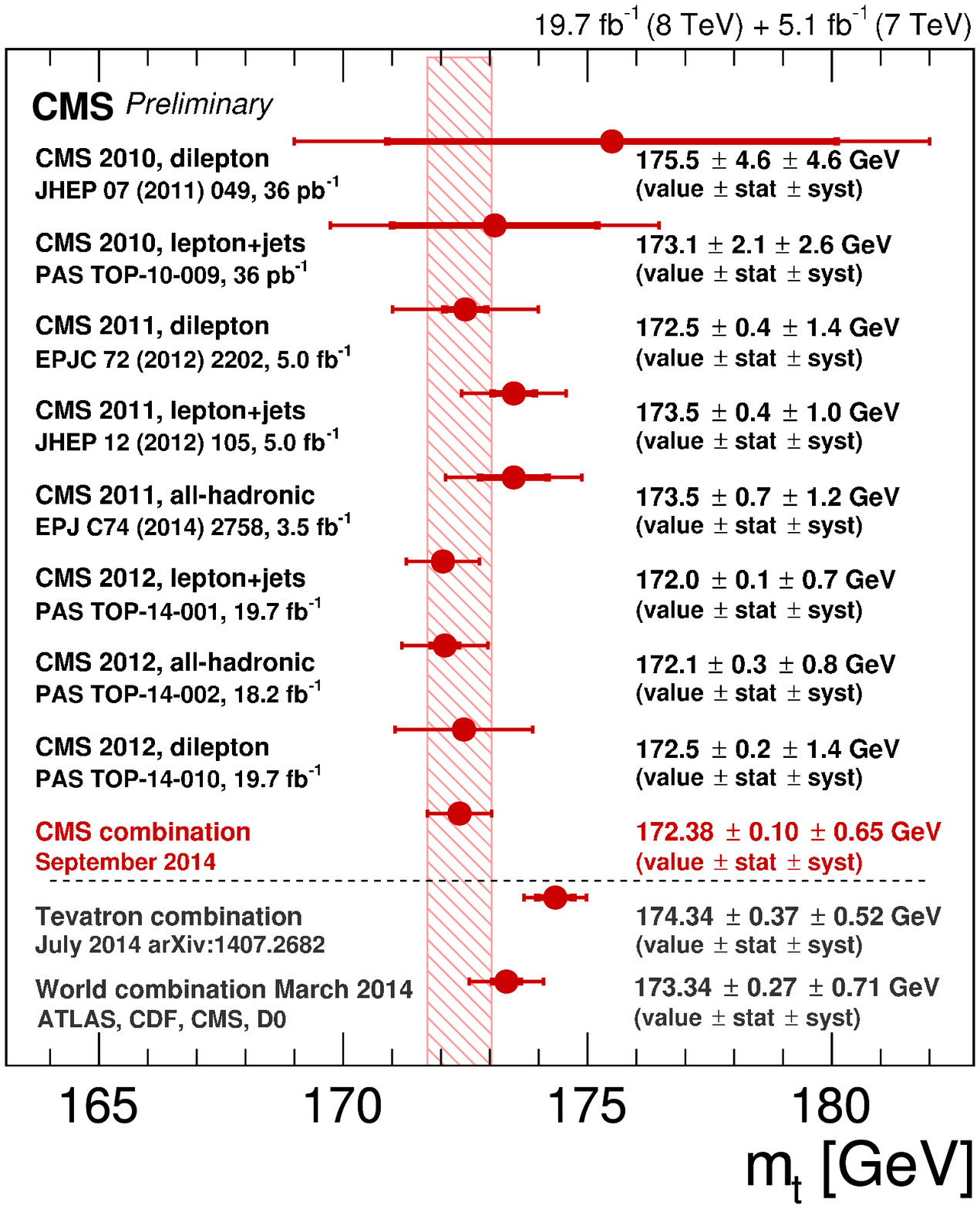}
\includegraphics[width=5cm]{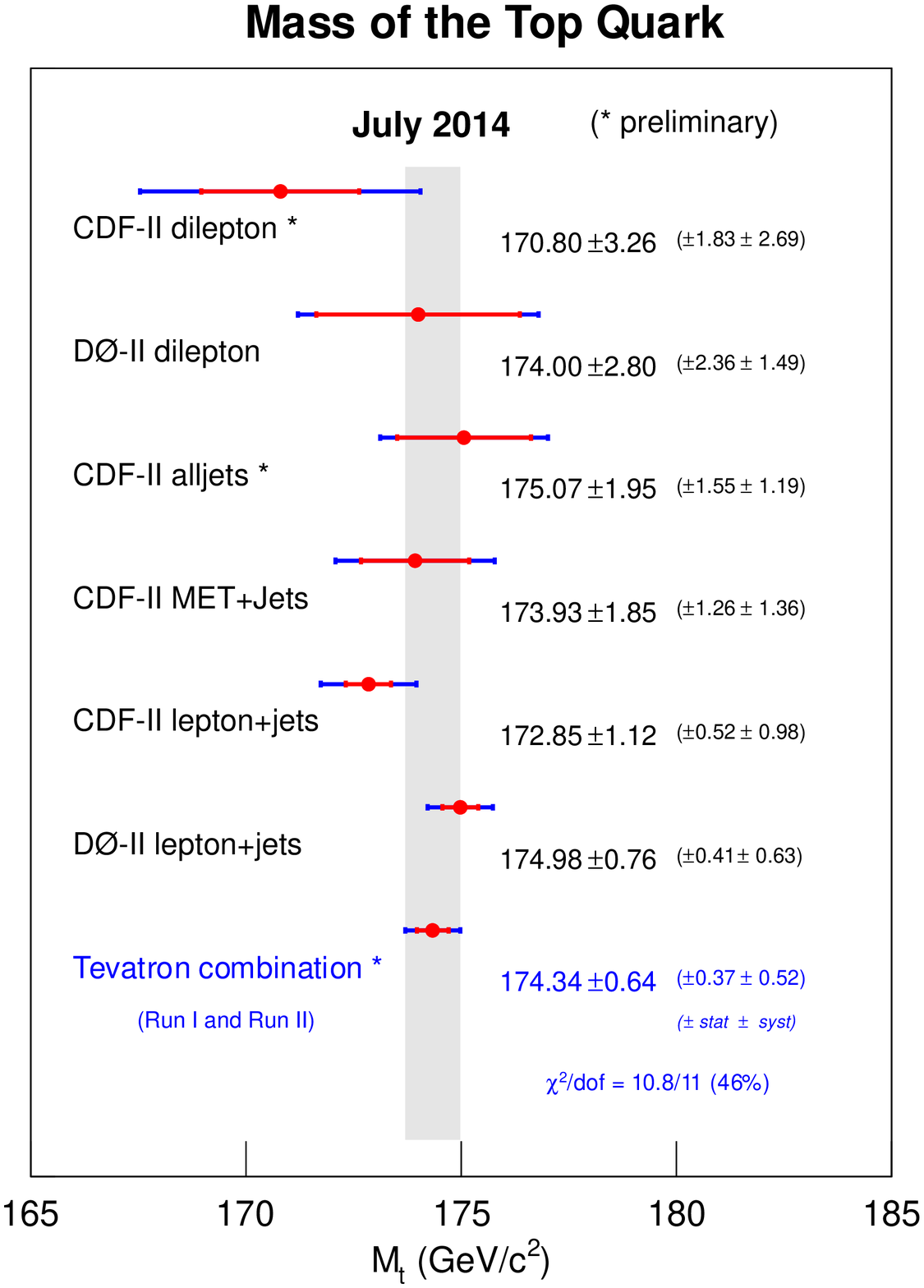}
\end{center}
\caption{{\bf Left}: Summary of the latest ATLAS direct $m_t$ measurements.
{\bf Center}: Summary of the eight CMS $m_t$ measurements and their combination.
{\bf Right}: CDF and D0 $m_t$ measurements and their combination.}
\label{fig11}
\end{figure}

\begin{figure}[ht]
\begin{center}
\includegraphics[width=6cm]{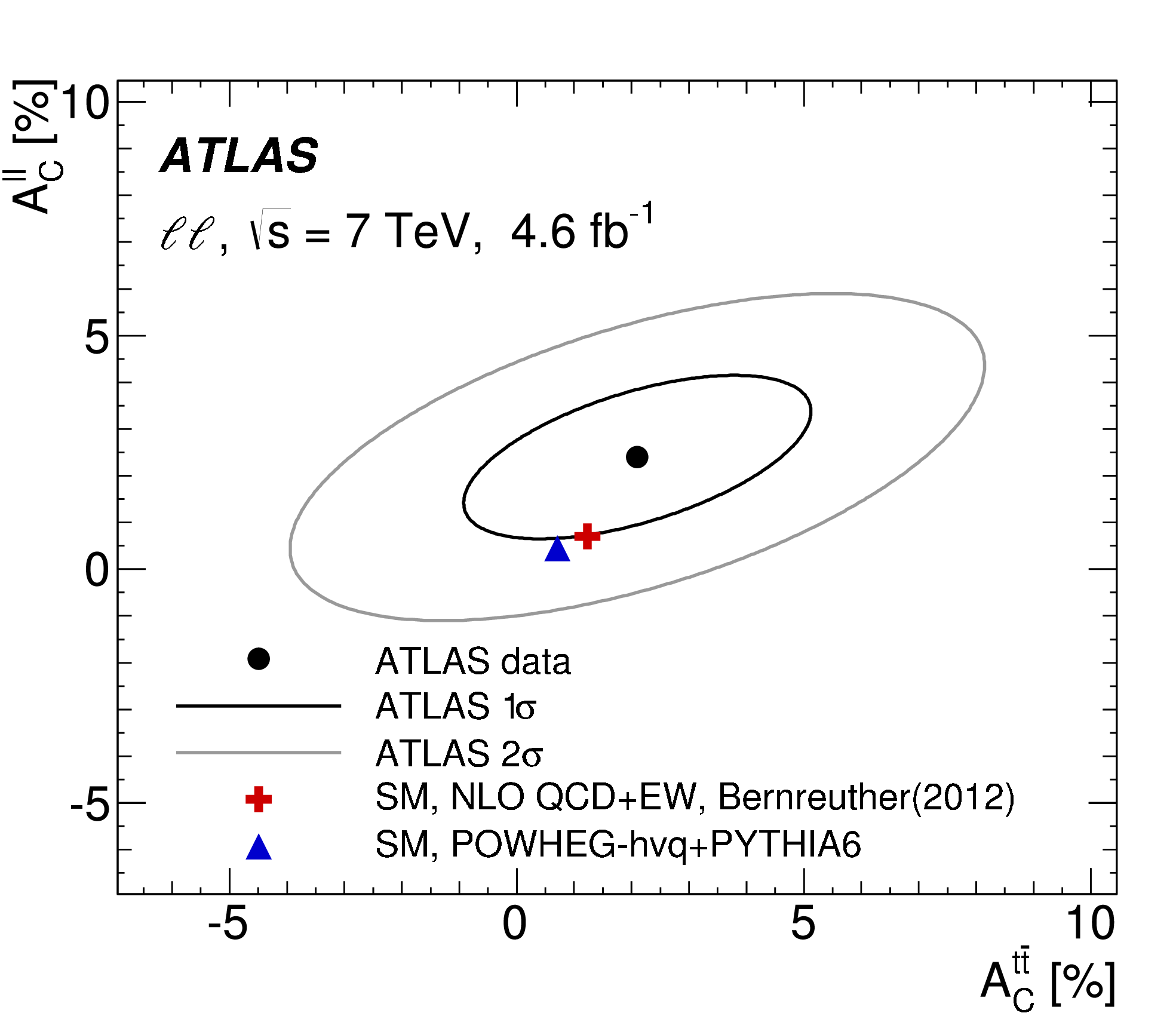}
\includegraphics[width=6cm]{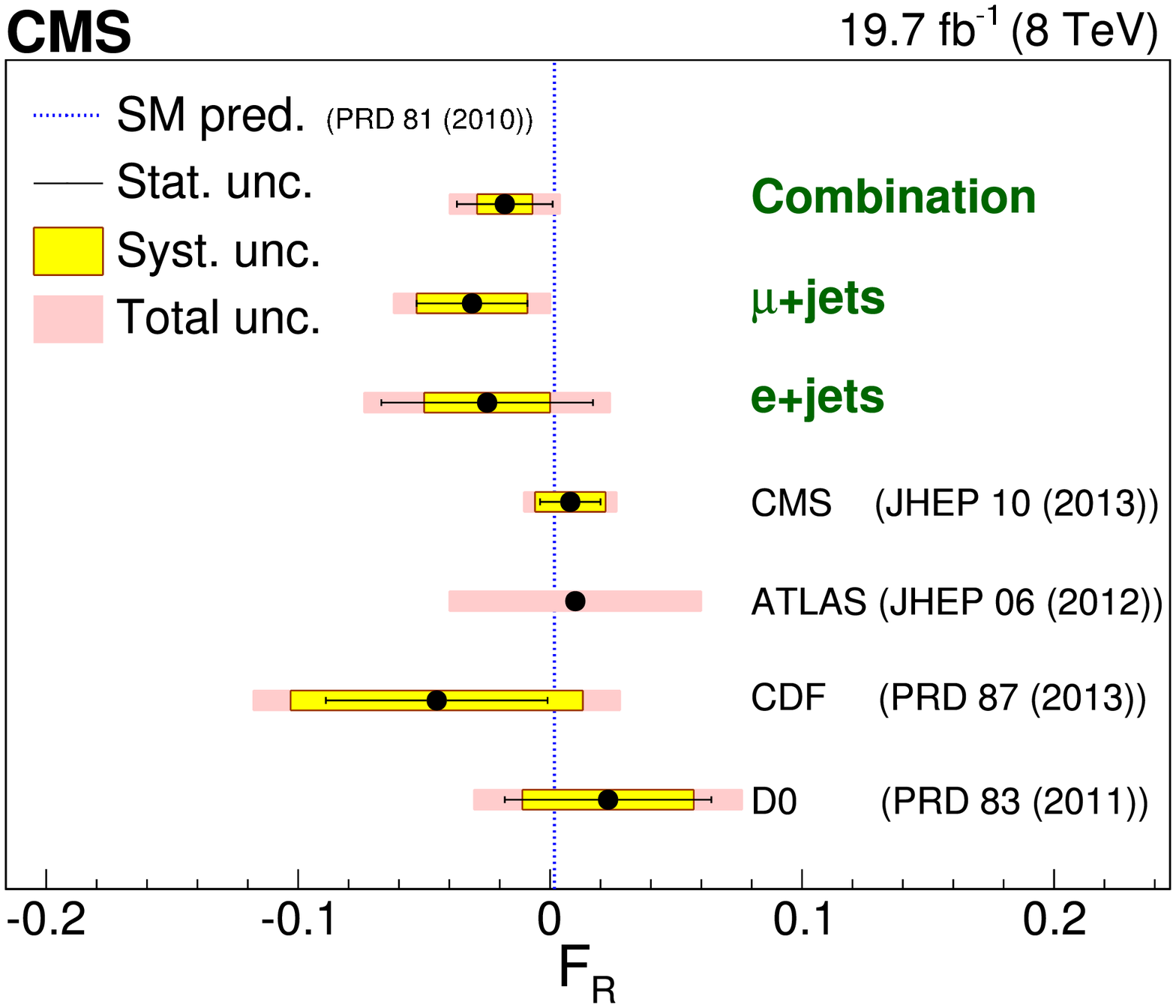}
\end{center}
\caption{{\bf Left}: Comparison of the inclusive $A_C^{ll}$ and $A_C^{t\bar{t}}$ measurement values to the theory predictions (SM NLO QCD+EW prediction).
{\bf Right}: The right-handed helicity fraction of the $W$ boson from the top quark decay.}
\label{fig12}
\end{figure}

\begin{figure}[ht]
\begin{center}
\includegraphics[width=5.cm]{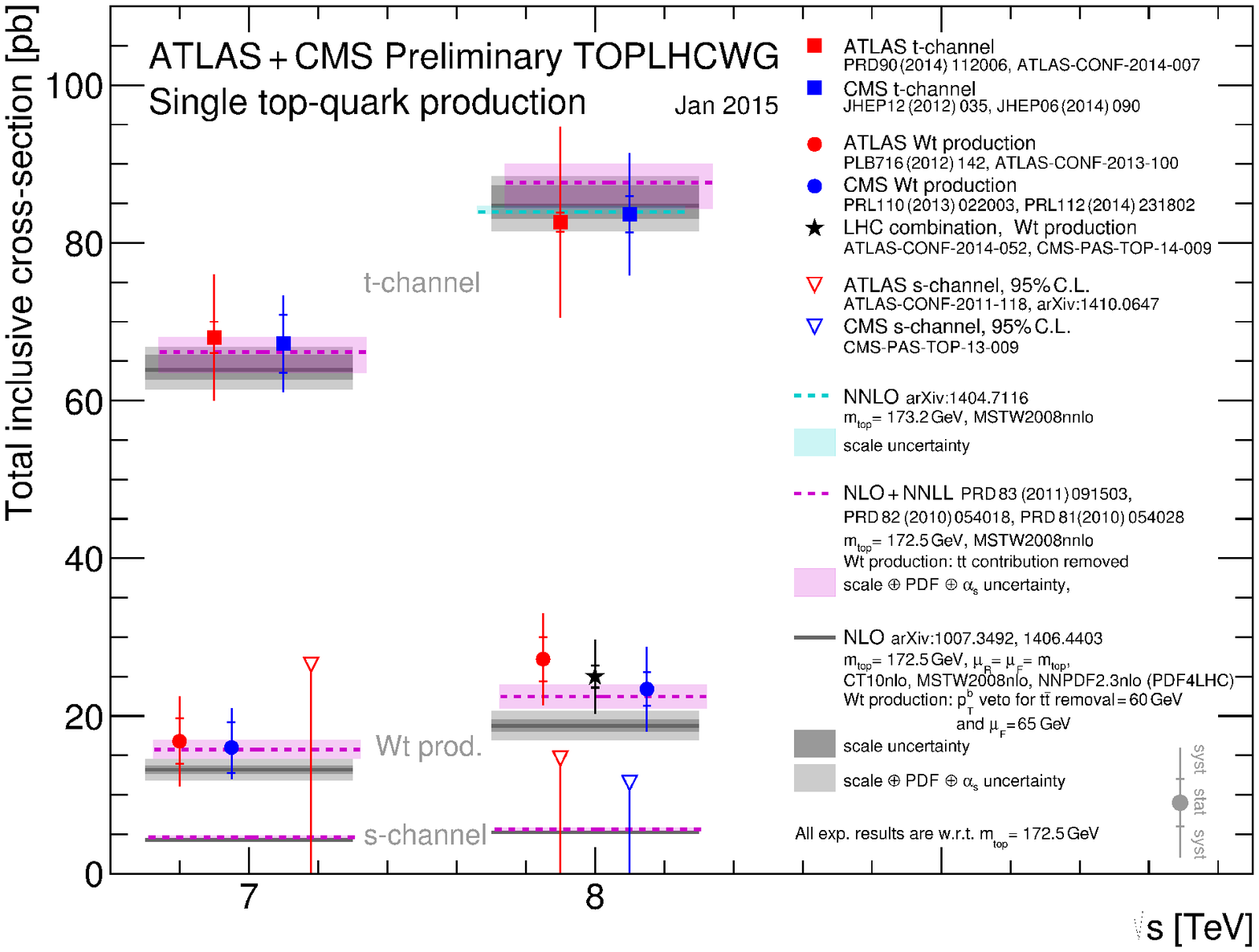}
\includegraphics[width=4.cm]{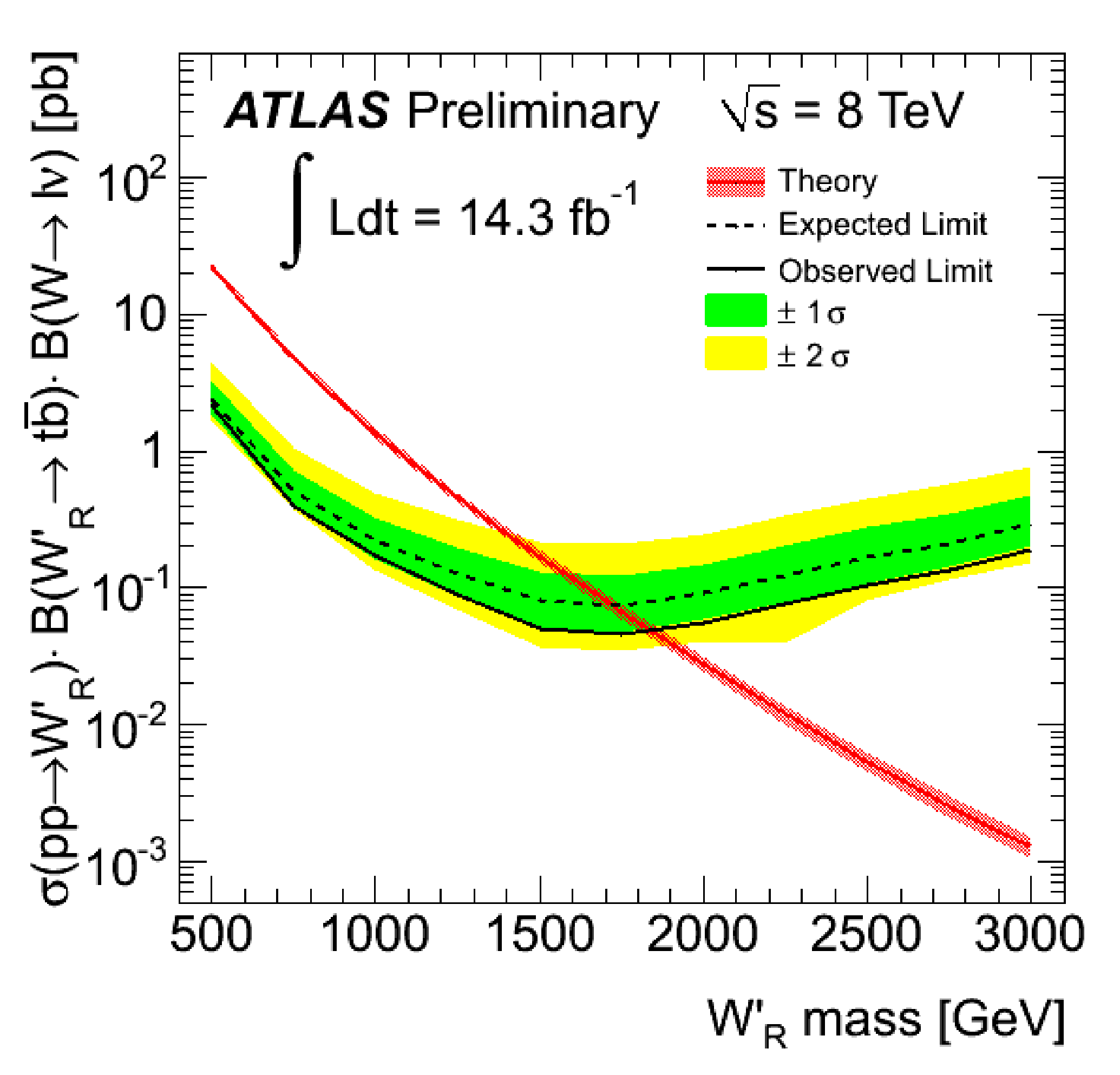}
\includegraphics[width=4.5cm]{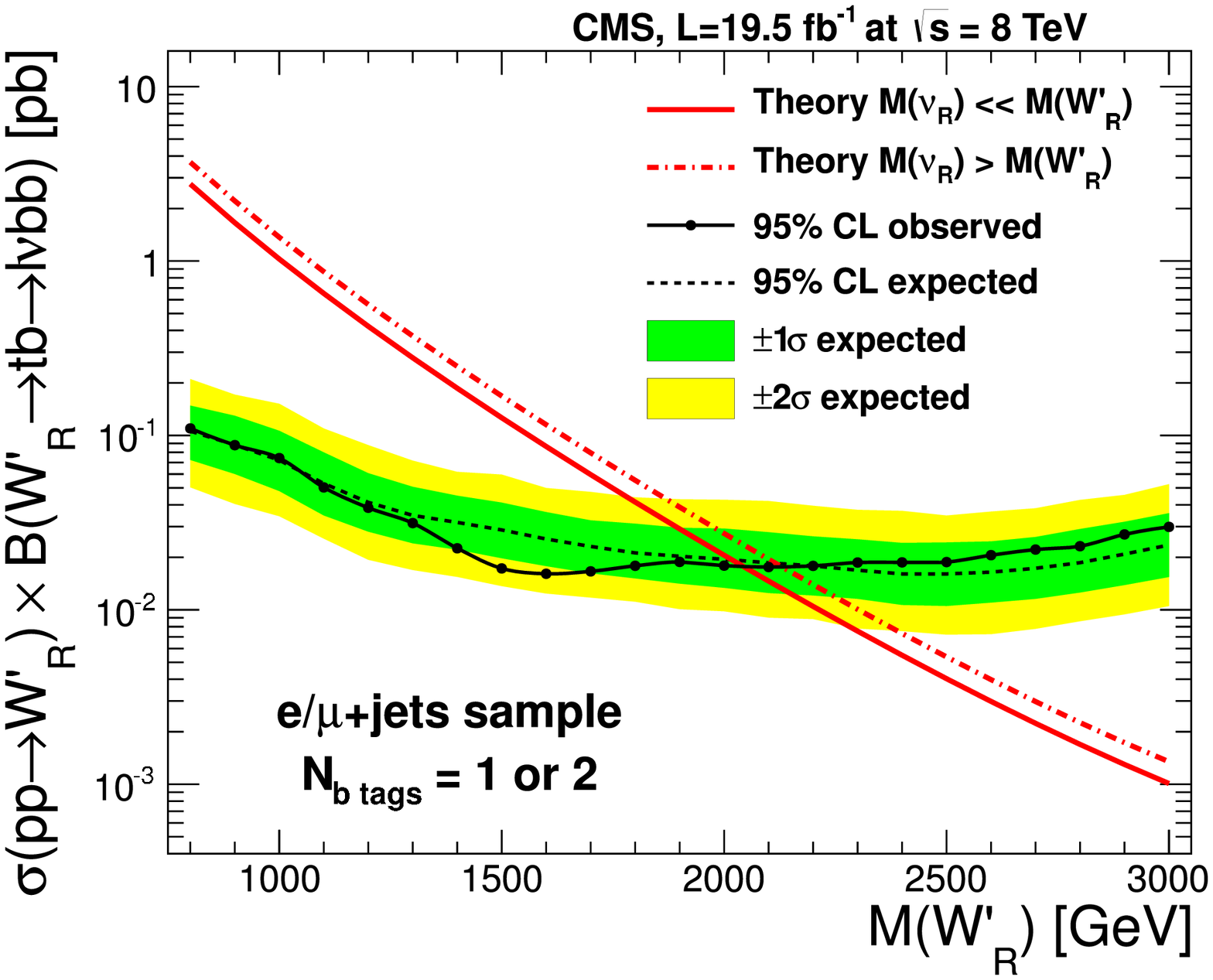}
\end{center}
\caption{{\bf Left}: Summary of ATLAS and CMS measurements of the single-top production cross-sections 
in various channels as a function of the center of mass energy.
{\bf Center}: Observed and expected 95\% CL limits on the $W'$-boson cross-section times 
branching ratio prediction, as a function of the mass of the $W'$ boson, for right-handed $W'$ bosons at ATLAS 8 TeV. 
{\bf Right}: Upper limits on the production cross-section of right-handed $W'$ 
bosons obtained for combination of the electron and muon channels at CMS 8 TeV.}
\label{fig13}
\end{figure}

J.G.~Garay, A.~Jung, S.~Protopopescu, and C.~Schwanenberger gave detailed presentations and overviewes on the current status of measurements
for top-quark pair production inclusive and differential cross sections at the ATLAS and CMS experiments.  
C.~Schwanenberger summarized results of the measurements of ${t\bar{t}}$ + jets, ${t\bar{t}} + \gamma$, ${t\bar{t}} + Z$, 
and ${t\bar{t}} + W$ production at the ATLAS experiment~\cite{Aad:2015uwa}.
Top-quark properties and decays were also extensively discussed.
Preliminary ATLAS and CMS combined results for the inclusive ${t\bar{t}}$ cross section measurements 
at $\sqrt{s} = 8$ TeV~\cite{ATLAS:2014aaa,CMS:2014gta} were presented. 
The main results are illustrated in Fig.\ref{fig9}(right), where LHC and Tevatron measurements of $t\bar{t}$ pair production 
cross section are shown as a function of the centre-of-mass energy and are compared 
to the NNLO QCD calculation complemented with NNLL resummation~\cite{Czakon:2013goa}. 
Examples of measurements of top-quark pair production differential cross sections 
at ATLAS~\cite{ATLAS:2014daa} and CMS~\cite{Khachatryan:2015oqa} at $\sqrt{s} = 8$ TeV, are shown in Fig.\ref{fig10}. 
Precise measurements for these cross sections are extremely important for global QCD 
analyses to constrain proton PDFs (in particular the gluon) in the large $x$ region, where these are currently poorly determined~\cite{Guzzi:2014wia}. 
Moreover, these measurements provide us with the possibility of pinning down the correlations between 
the cross section, PDFs, $\alpha_s$, and top-quark mass. 
S.~Menke and H.~Liu presented measurements of the top-quark mass $m_t$ at the LHC and Tevatron respectively.
Very recent $m_t$ measurements at ATLAS~\cite{Aad:2015nba} 7 TeV and CMS~\cite{CMS:2014hta} at 8 TeV are shown in Fig.\ref{fig11}, where
the CMS summary result of LHC Run-I is $m_t = 172.38 \pm 0.10(stat) \pm 0.65(sys)$ GeV. 
The preliminary ATLAS summary result of the 7 TeV Run-I is $m_t = 172.99 \pm 0.48(stat) \pm 0.78(sys)$ GeV.
The March 2014 Tevatron plus LHC combined result is $m_t = 173.34 \pm 0.27(stat) \pm 0.71(sys)$ GeV. 
The results presented by H.~Liu for the $m_t$ combined measurements from CDF and D0 experiments at the Tevatron collider~\cite{Tevatron:2014cka} 
are shown in Fig.\ref{fig11}(right). The Tevatron combination resulted in $m_t = 174.34 \pm 0.37 (stat) \pm 0.52 (syst)$ GeV. 
Measurements of the top-quark properties in the production and decays of $t\bar{t}$ events at 
CMS~\cite{Khachatryan:2014vma} and ATLAS~\cite{Aad:2015jfa} were discussed by A.~Jung.
In particular he discussed measurements of the top-quark pair charge asymmetry, $W$ helicity in top decays, top-quark charge, 
$t\bar{t}$ spin correlation, and searches for anomalous couplings. 
A sample of the main results is shown in Fig.\ref{fig12}(left) (ATLAS) and in Fig.\ref{fig12}(right) (CMS), where they are 
compared to the SM predictions and are found to be in agreement. 
K.~Finelli presented measurements of single-top production cross section at ATLAS~\cite{ATLAS:2014dja} and CMS~\cite{Khachatryan:2014iya}.
Inclusive and differential cross sections measured in all channels were discussed.
The ATLAS and CMS preliminary combination for the total inclusive cross section for single top production is shown in Fig.\ref{fig13}(left).
Single-top physics is a probe for BSM physics. In fact, single top-quark production cross sections set stringent limits on the mass of extra charged currents 
decays of $W'\rightarrow tb$. Two recent studies are shown at ATLAS~\cite{TheATLAScollaboration:2013iha} 
and CMS~\cite{Chatrchyan:2014koa} in Fig.\ref{fig13}(center, right), where 
exclusion limits for the mass of right-handed $W'$s are given. 

\section{Heavy flavours in global QCD analyses of proton PDFs}

This section is concerned with the impact of heavy flavours in the determination 
of the structure of the proton in global QCD analyses of world experimental data. 
Measurements of heavy-flavour cross sections play a crucial role in the most 
recent PDF global analyses~\cite{Dulat:2015mca,Harland-Lang:2014zoa,Ball:2014uwa,Abramowicz:2015mha,Alekhin:2013nda,Jimenez-Delgado:2014twa,Owens:2012bv}, 
and are an essential ingredient of the LHC run-II program~\cite{Rojo:2015acz}. 
LHC run-II measurements will set a completely new frontier for PDFs accuracy, as they will be constrained 
in new kinematic regions. Precise determination of PDFs and reduction of their uncertainties are crucial 
for a correct estimate and characterization of LHC cross sections and for new physics searches.   
In the studies presented at this workshop, several very recent heavy-flavour cross section measurements have been exploited 
to constrain PDFs in kinematic regions so far unexplored. 

\begin{figure}[ht]
\begin{center}
\includegraphics[width=5cm]{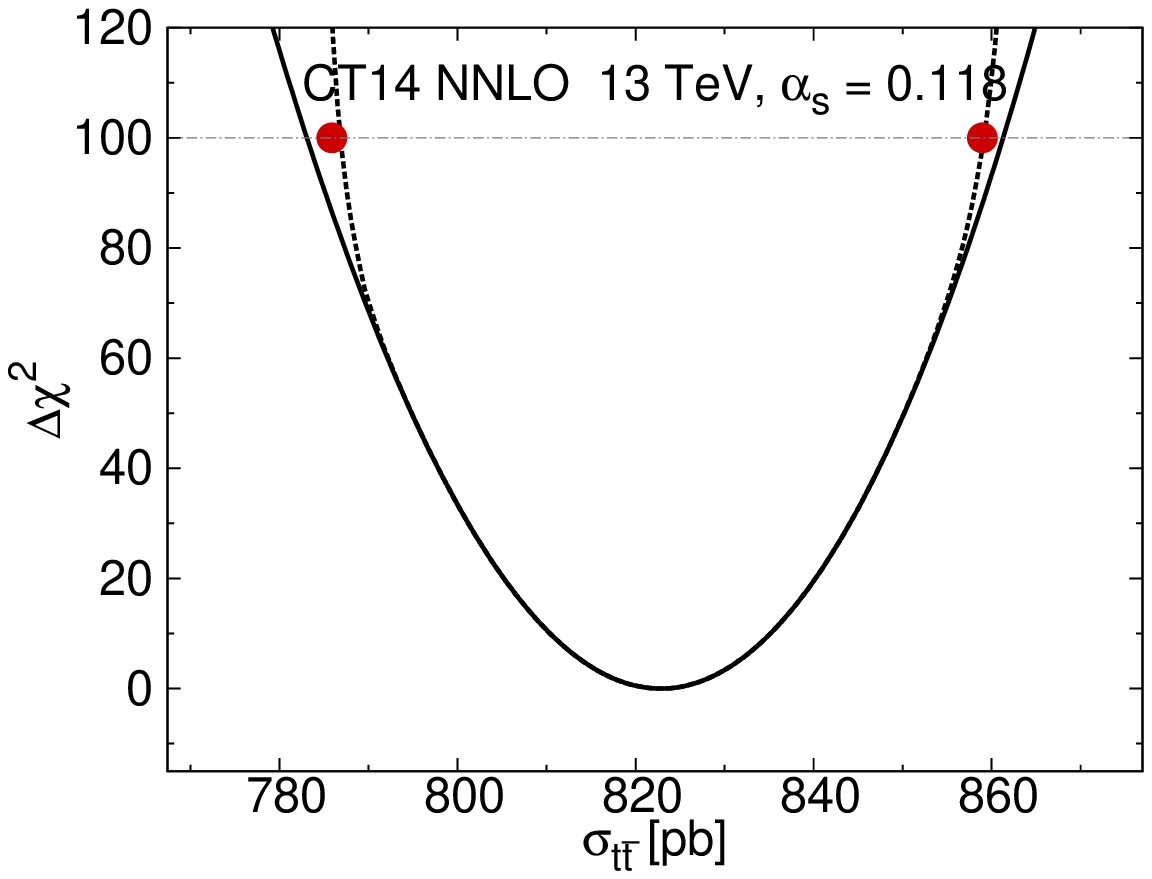}
\includegraphics[width=5cm]{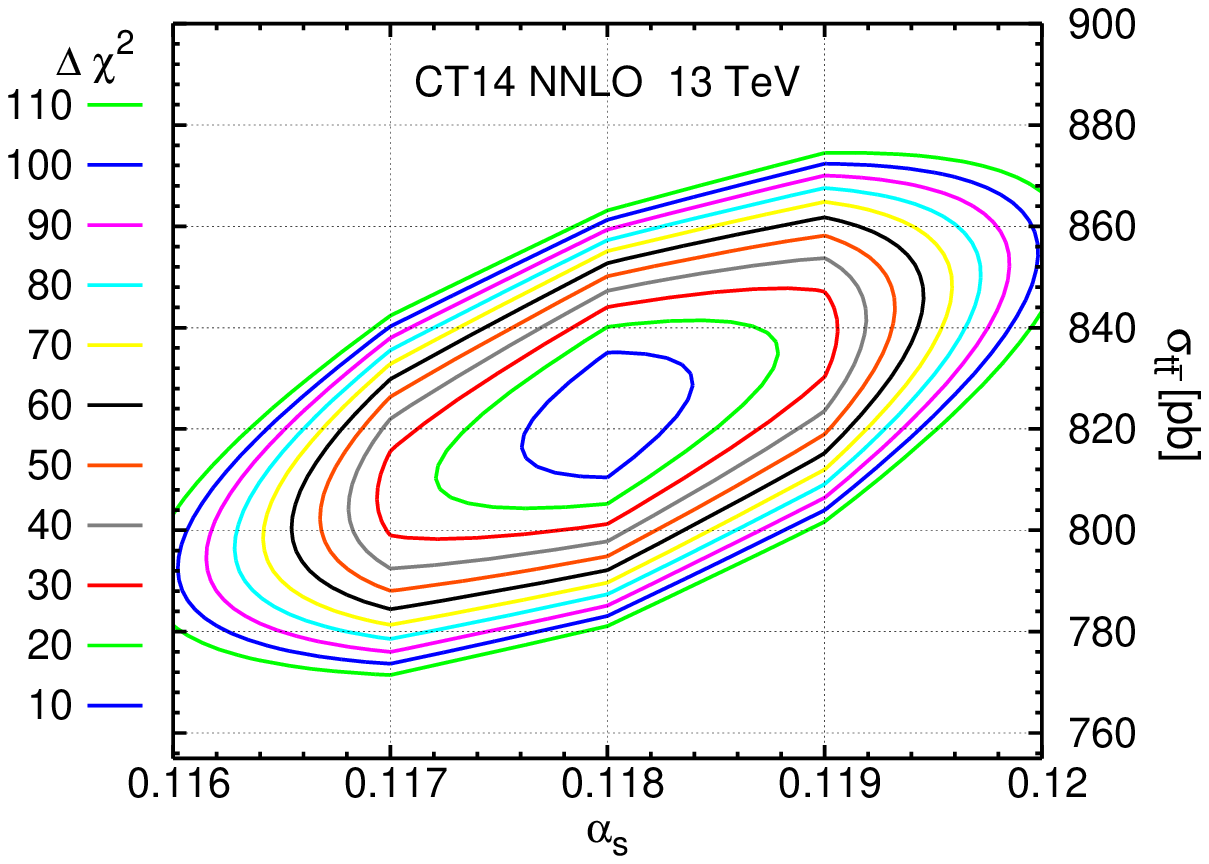}
\end{center}
\caption{{\bf Left}: Dependence of the increase in the constrained CT14 fit 
on the expected cross section at the LHC 13 TeV, for $\alpha_s(M_Z)$ = 0.118. 
{\bf Right}: Contour plot of $\Delta\chi^2(\sigma_{t\bar{t}}, \alpha_s(M_Z))$ in the $(\sigma_{t\bar{t}}, \alpha_s(M_Z))$ plane at the LHC 13 TeV.}
\label{fig14}
\end{figure}

T.J.~ Hou presented a detailed analyses of the predictions for $gg \rightarrow H$ and $t\bar{t}$ cross sections at the LHC 7, 8 and 13 TeV,
as well as their uncertainties from both the PDFs and the strong coupling $\alpha_s$, in the context of the recent 
CT14 global QCD analysis~\cite{Dulat:2015mca}. In particular, he discussed a consistency check between the Lagrange multiplier and Hessian 
method. Some of the main results are shown in Fig.\ref{fig14}, where the behaviour of the $\chi^2$ as a function of 
the total inclusive NNLO $t\bar{t}$ pair production cross section (left), 
and a contour plot (right) of $\Delta\chi^2(\sigma_{t\bar{t}}, \alpha_s(M_Z))$ in the $(\sigma_{t\bar{t}}, \alpha_s(M_Z))$ plane at the LHC 13 TeV, are illustrated.
This type of test is extremely important to validate the hessian analysis 
and check the consistency of PDF uncertainties
in kinematic regions where there are no data to set constraints.

\begin{figure}[ht]
\begin{center}
\includegraphics[width=5cm]{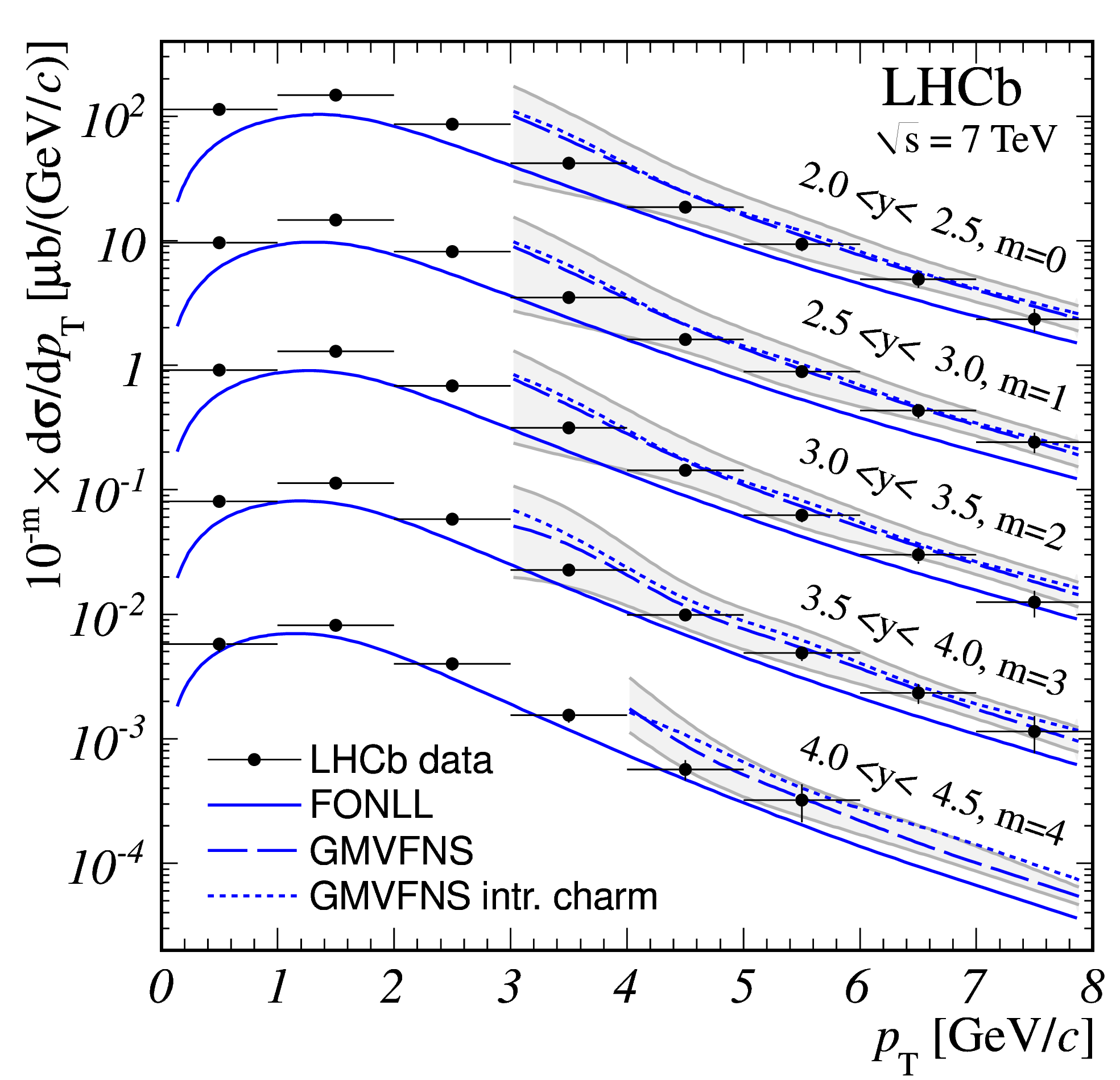}
\includegraphics[width=5cm]{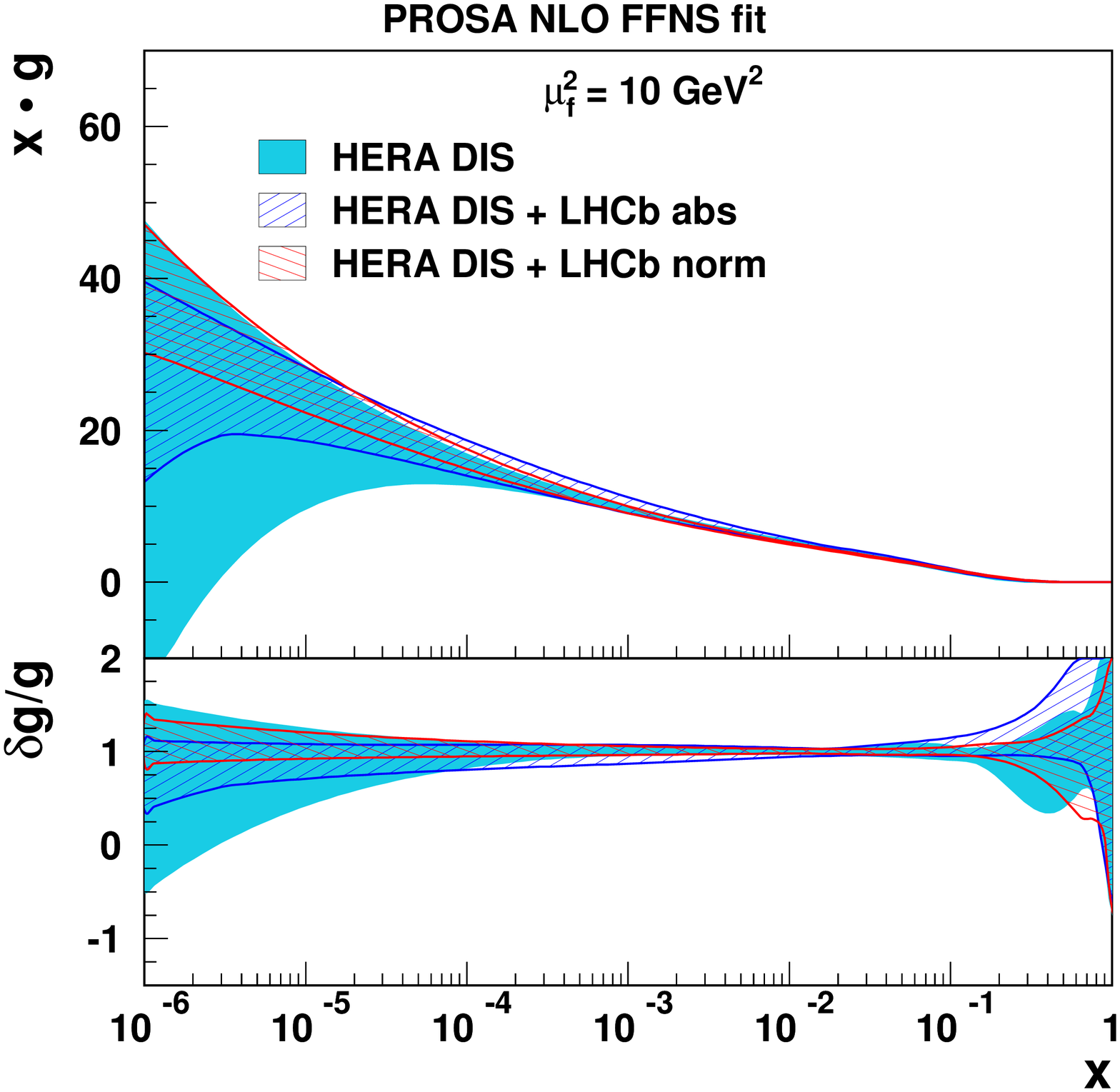}
\end{center}
\caption{{\bf Left}: Differential cross-sections for the charmed D0 meson production at the LHCb experiment, compared to NLO theoretical predictions.   
{\bf Right}: Gluon PDF uncertainty as obtained from the QCD fit. }
\label{fig15}
\end{figure}

A. Geiser presented a novel analysis~\cite{Zenaiev:2015rfa} on behalf of the Prosa collaboration~\cite{prosa}, 
in which recent LHCb measurements of charm and beauty production cross sections~\cite{Aaij:2013mga,Aaij:2013noa}, 
in conjunction with HERA data~\cite{Aaron:2009aa}, are used in a NLO PDF analysis to directly constrain the gluon PDF down to $x \approx 5\cdot 10^{-6}$.
This kinematic range is currently not covered by other experimental data in perturbative QCD fits.
One of the key differential cross-section measurements for the charmed D0 meson production 
at the LHCb experiment, is illustrated in Fig.\ref{fig15}(left).
The main finding of this analysis is illustrated in Fig.\ref{fig15}(right), 
where the gluon PDF is found to be positive and well constrained at $x\approx 10^{-6}$.

\begin{figure}[ht]
\begin{center}
\includegraphics[width=6.5cm]{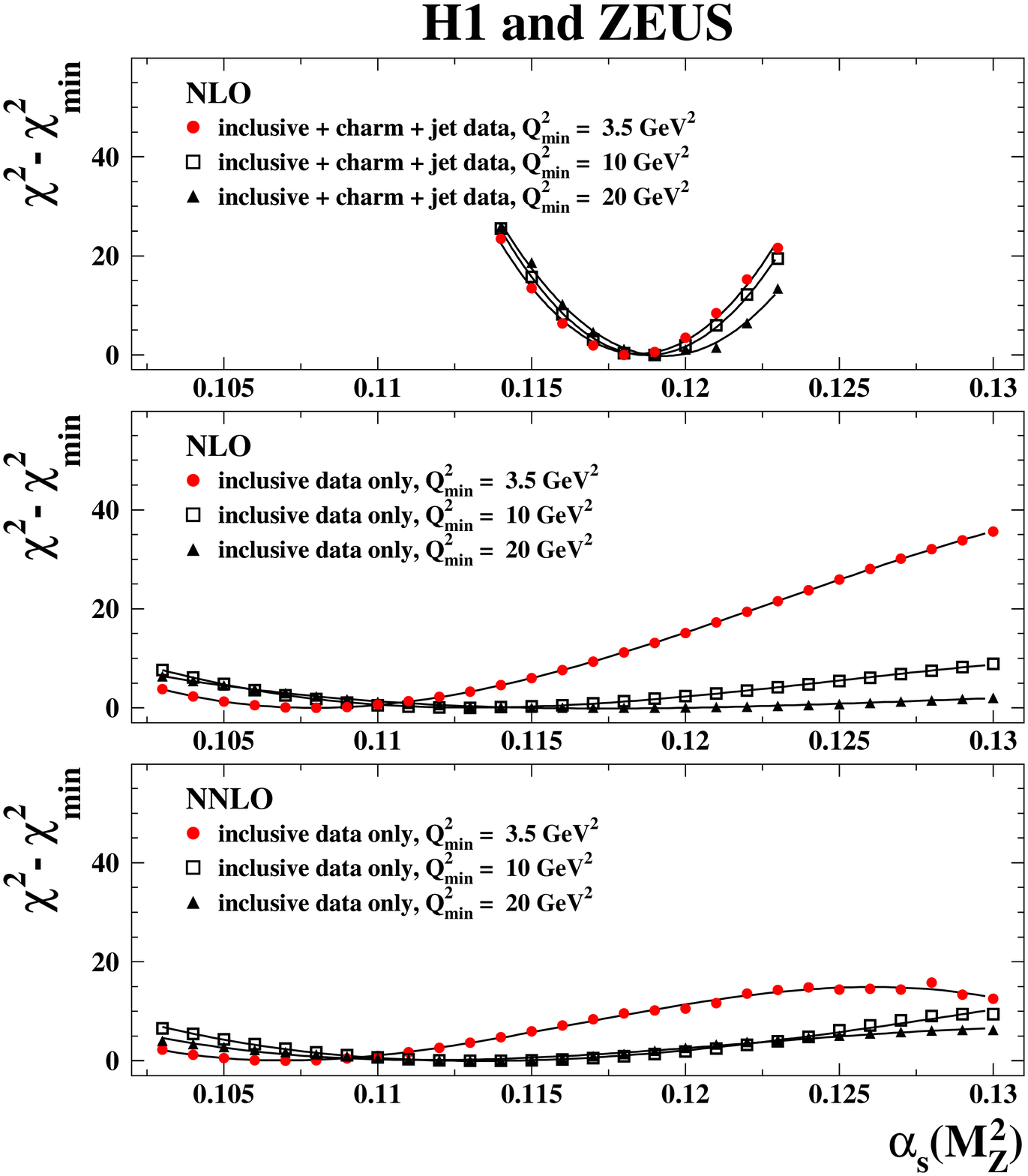}
\includegraphics[width=6.5cm]{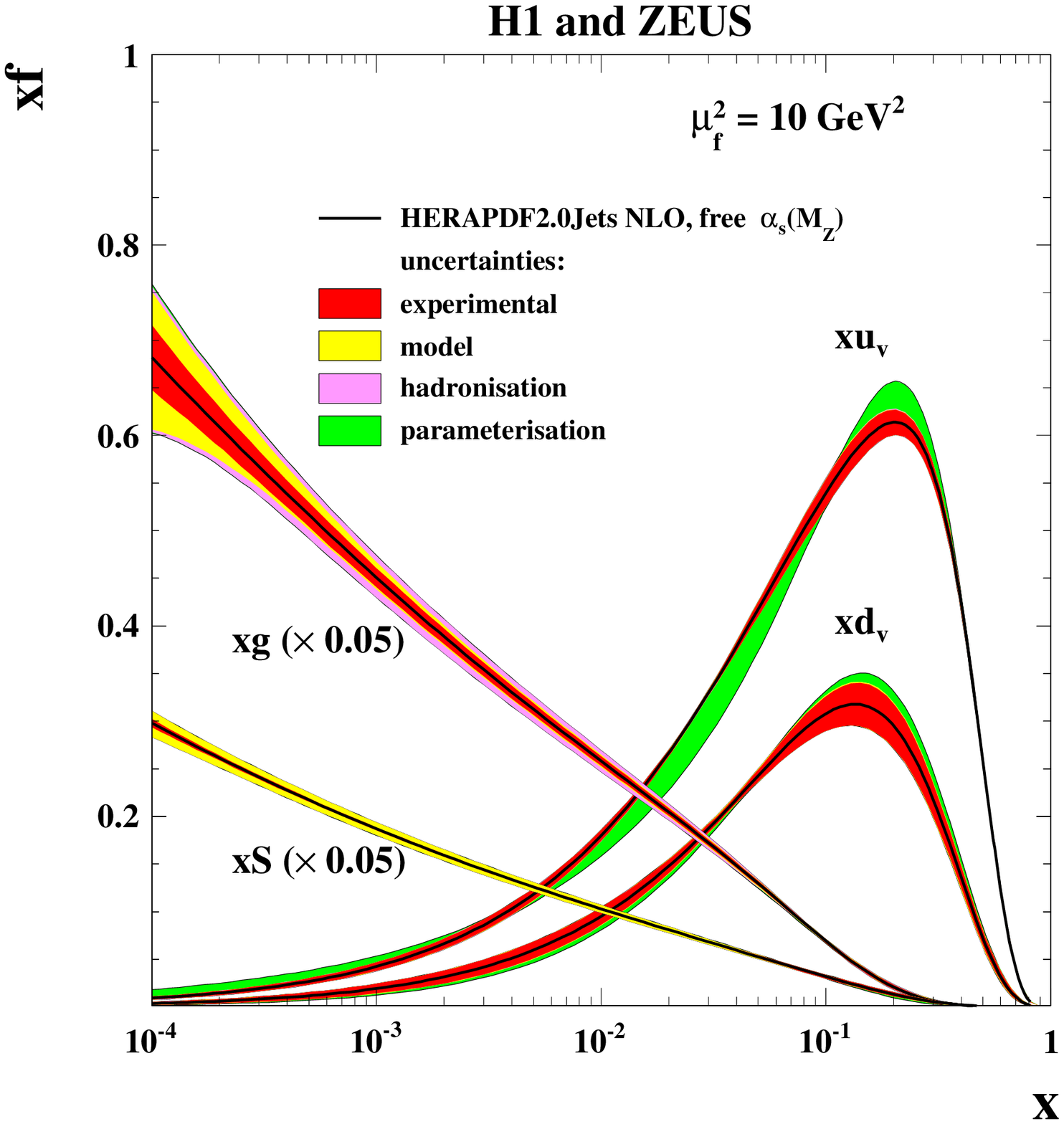}
\end{center}
\caption{{\bf Left}: $\Delta \chi^2 = \chi^2 - \chi^2_{\rm min}$ vs. $\alpha_s(M^2_Z)$ for pQCD fits 
with different $Q^2_{\rm min}$ using data on (a) inclusive, charm and jet production at NLO,
(b) inclusive $ep$ scattering only at NLO and (c) inclusive $ep$ scattering only at NNLO.
{\bf Right}: The parton distribution functions $x u_v$, $x dv$, $x S = 2x( \overline{U} + \overline{D})$ 
and $xg$ of HERAPDF2.0Jets NLO at $\mu_f^2= 10$ GeV$^2$ with free $\alpha_s(M^2_Z)$.}
\label{fig16}
\end{figure}

G.~Brandt discussed an extension of the HERAPDF2.0 QCD fit~\cite{Abramowicz:2015mha} in which 
data of charm and jet-production cross section are used to obtain a simultaneous
determination of PDFs and $\alpha_s(M^2_Z)$. The new PDFs obtained in this simoultaneous fit are called HERA2.0Jets~\cite{Abramowicz:2015mha}. 
The resulting value of the strong coupling constant is 
$\alpha_s(M^2_Z) = 0.1183 \pm 0.0009(exp) \pm 0.0005(model/param.) \pm 0.0012(hadronisation) + 0.0037 - 0.0030(scale)$. 
This value is in excellent agreement with the world averge $\alpha_s(M^2_Z) = 0.1185$.
In Fig.\ref{fig16}(left) shown are different plots of the resulting 
$\chi^2$ for various determinations of $\alpha_s(M^2_Z)$.
In the (right) panel the HERAPDF2.0Jets PDFs at factorization scale $\mu_f^2= 10$ GeV$^2$ obtained from a NLO fit with free $\alpha_s(M^2_Z)$, 
are illustrated. A significant reduction of the uncertainty of the gluon PDF is found.

\section{Conclusions}
Heavy-flavour physics has many implications on the multifaceted aspects of hadronic matter.
It is a powerful tool to probe features of QCD in different contexts 
and it is crucial for searches of new physics. 
At this conference, a large number of very interesting new results were presented
confirming that, even though an impressive amount of important 
results have already been achieved, 
a lot of work is ahead to explore key features 
of QCD and electroweak theory which are crucial to face the challenges of LHC run-II.  

\section{Acknowledgments}

This work is supported in part by the Lancaster-Manchester-Sheffield 
Consortium for Fundamental Physics under STFC grant ST/L000520/1.



\end{document}